# Exciton Assisted Deeply Subwavelength Nano-Photonics

Haonan Ling[1], Arnab Manna[3], Jialiang Shen[4], Ho-Ting Tung[1], David Sharp[3], Johannes Fröch[2,3], Siyuan Dai[4,*], Arka Majumdar[2,3,*], Artur R. Davoyan[1,*]

[1]Department of Mechanical and Aerospace Engineering, University of California, Los Angeles, CA, USA
[2]Department of Electrical and Computer Engineering, University of Washington, Seattle, WA, 98195, USA
[3]Department of Physics, University of Washington, Seattle, WA, 98195, USA
[4]Materials Research and Education Center, Department of Mechanical Engineering, Auburn University, Auburn, AL, 36849, USA

**Abstract:**
The wave nature of light sets a fundamental diffraction limit that challenges confinement and control of light in nanoscale structures with dimensions significantly smaller than the wavelength. Here, we demonstrate van der Waals $MoS_2$ nano-photonic devices with dimensions as small as $\simeq \lambda/16$ (~60 nm at 1000 nm excitation wavelength). This deep subwavelength light confinement is achieved by exploiting the coupling between $MoS_2$ excitons and photons. We validate deep subwavelength light control via far- and near-field measurements. Our near-field measurements reveal detailed imaging of excitation, evolution, and guidance of fields in $MoS_2$ nanodevices, whereas our far-field study examines highly confined integrated photonics. Exciton-driven nano-photonics at a fraction of a wavelength demonstrated here could dramatically reduce the size of integrated photonic devices and opto-electronic circuits with potential applications in optical information science and engineering.

**Main text:**
Subwavelength light coupling and manipulation [1-3] hold great promise for a wide range of science and engineering disciplines. Nano-photonics has already influenced exascale and quantum computing [4-8], photonic integrated circuits for ultra-broadband communications [9-12], efficient bio and chemical sensing [13-16], and improving climate sustainability [17, 18] among others. By further bridging the gap between optical fields ($\lambda \sim 1\mu m$) and much smaller intrinsic material excitations, typically manifested at $\ll 100\ nm$, the full potential of nano-photonics can be harnessed. Yet, coupling, guidance and control of light in structures with deeply subwavelength features $< 100\ nm$ ( $< \lambda/10$ ), to date, remains a challenge. Here, we demonstrate that electronically bulk molybdenum disulfide ($MoS_2$) – a layered van der Waals semiconductor in the family of transition metal dichalcogenides (TMDCs) [19-21] – enables integrated photonic devices as thin as $\simeq \lambda/16$ (figure 1a). With the use of scattering near-field microscopy and far-field imaging we examine properties of light guidance and coupling within a range of deeply subwavelength devices. The observed light control at the fraction of a wavelength is attributed to a unique $MoS_2$ lightwave dispersion driven by intrinsic coupling between excitons and photons.

Presently, subwavelength ($<\lambda/2$) photonics is based on two key material avenues – dielectrics and metals. All-dielectric systems, featuring low optical extinction below the bandgap ($n \gg k$ and $k \to 0$, where $n$ is refractive index and $k$ is extinction coefficient), are very efficient at controlling light at the scale $\simeq \frac{\lambda}{2n}$. Such areas as photonic integrated circuits, metasurface optics, and on-chip opto-electronics have benefited from all-dielectric photonic devices [5, 22-28]. However, the limited range of refractive indices [29, 30] of conventional dielectrics and semiconductors (typically $n < 3.5$ in the near-infrared band; see Supplementary Section I) challenges creating deeply subwavelength integrated photonic devices. Metals offer an alternative, where strong coupling and hybridization with bulk charge density waves drastically modify lightwave dispersion, giving rise to highly confined surface plasmon-polaritons with a high effective guide index [1,31]. The ability of these excited states to squeeze light well below the diffraction limit found a wide range of applications, from interconnects to quantum optics to photochemistry [9, 10, 18, 32]. Nonetheless, high free electron density in metals inevitably contributes to optical loss [33, 34], which narrows the range of applications for plasmonic components and devices (Supplementary Section I).

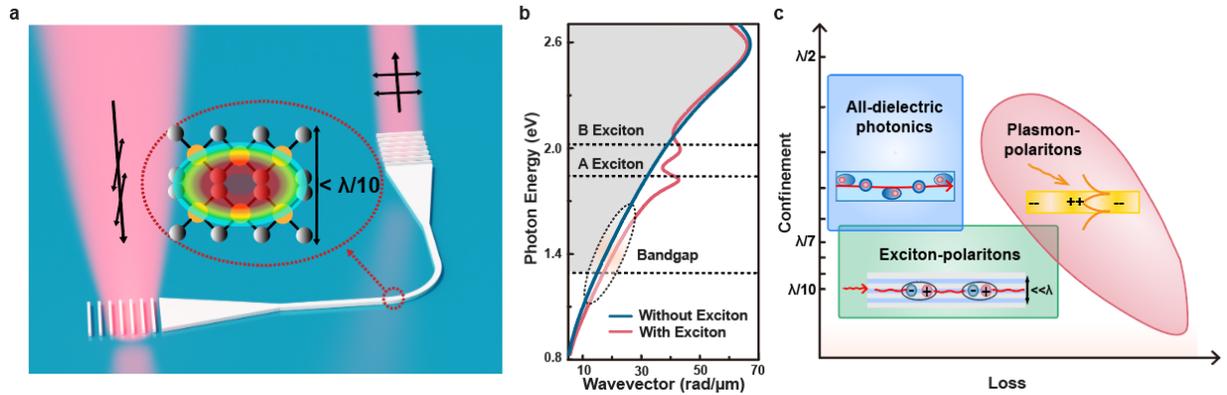

*Figure 1. Deeply subwavelength photonics mediated by excitons. (a) Conceptual illustration of a MoS$_2$ integrated photonic waveguide studied in this work. Light can be coupled, confined and guided in devices thinner than $< \lambda/10$. (b) Guided wave dispersion in a 60 nm thick MoS$_2$ waveguide with and without exciton contribution. Shaded area corresponds to bulk band gap where optical absorption is high. The highlighted region indicates wavelength range studied here. (c) Conceptual comparison between different material platforms. Similar to plasmon-polaritons, exciton driven photonics can enable deeply subwavelength devices, while offering low below-bandgap extinction.*

Light coupling to intrinsic material excitations is also naturally attained in van der Waals transition metal dichalcogenides (MX$_2$ compounds where M = Mo, W; X = S, Se, Te), in which excitons with large binding energies are excited [19-21,35]. Interaction of photon modes with excitons in electronically bulk TMDCs leads to hybrid waves [36-38] (extensive study and derivation of exciton–photon coupling and hybridization is provided in Supplementary Section I). Strong exciton-photon interaction and formation of exciton-polaritons have been examined in a number of bulk transition metal dichalcogenide structures [39-44]. Importantly, similar to plasmonics [31], exciton—photon interaction yields a strongly modified wave dispersion, which contributes to a high guided index well below the exciton resonance (figure 1b). However, unlike metals suffering from free electron damping [33,34], below bandgap absorption in excitonic materials is expected to be low [45-49] (figure 1b, see also Supplementary figures S3 and S7), giving an opportunity to create efficient integrated photonic devices [45, 49].

The utility of high refractive index bulk TMDCs has already been recognized in the design of nanoscale resonators [40, 41, 50-52] and metasurfaces [53, 54]. Near-field measurements further revealed that hybrid exciton-polaritons could be excited and guided in exfoliated flakes of transition metal dichalcogenides [39, 55]. At the same time, light coupling to integrated photonic devices made of TMDCs and wave dispersion within subwavelength nanostructures are yet to be explored. Here, we employ scattering near-field microscopy and far-field imaging spectroscopy to examine a range of passive integrated MoS$_2$ photonic devices, including ridge waveguides, beam splitters, and interferometers. We study the performance of these devices in the vicinity of the exciton resonance and across the edge of the bulk MoS$_2$ bandgap (750 nm – 1050 nm), as marked in figure 1b. We demonstrate that by utilizing strong exciton-driven dispersion in MoS$_2$ deeply subwavelength, i.e. $< \lambda/16$, integrated photonic devices with small mode volume and device footprint (figure 1a) can be created. Deeply subwavelength light confinement in MoS$_2$, and other excitonic TMDCs [21], expands the nanophotonic pallet, figure 1c.

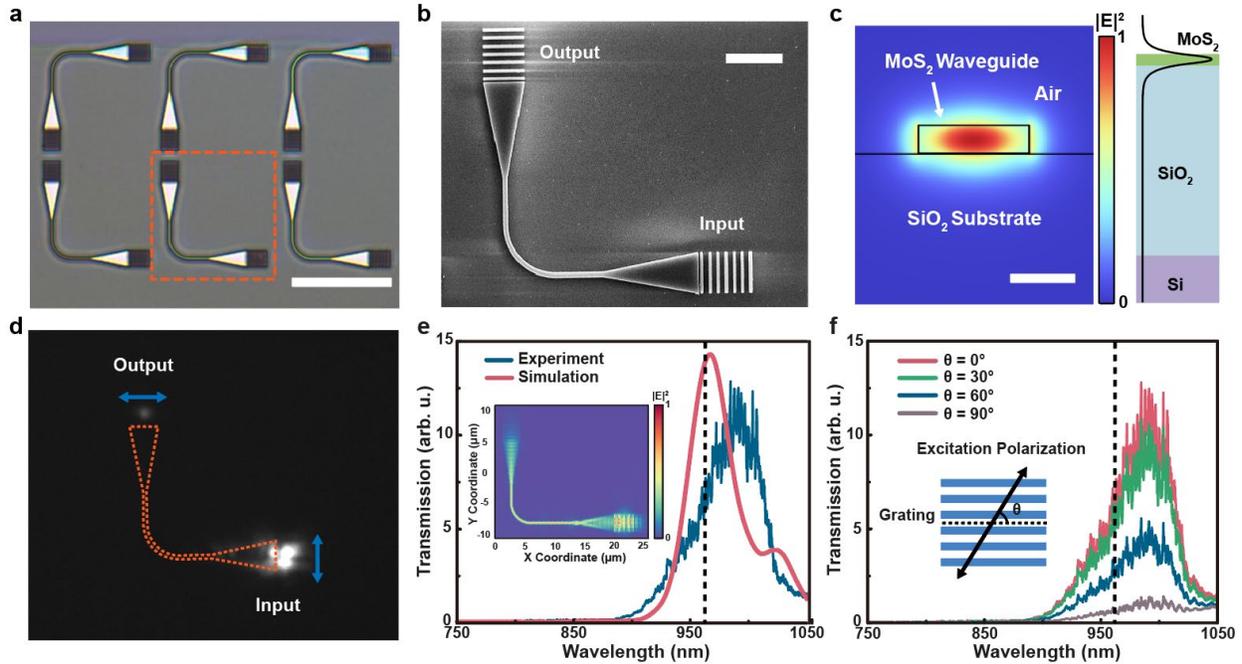

*Figure 2. Far-field study of deeply subwavelength MoS$_2$ waveguides.* *(a) Optical micrograph of fabricated waveguides. Scale bar: 20 μm. (b) SEM image of the fabricated device. Scale bar: 4 μm. (c) Optical mode profile (left) and a related cross-section (right) at 1 μm wavelength. The height and width of the device are 85 nm and 320 nm, respectively. (d) Far-field image of the scattered light collected in cross polarized light with a contour of the device outlined by a dashed curve. The arrows indicate the polarizations of input and output beam paths. (e) Measured and simulated waveguide transmission spectra. Inset shows simulated field profile at 1 μm. (f) Polarization dependent waveguide transmission spectra. In (e) and (f) dashed lines denote the wavelength of bulk MoS$_2$ bandgap.*

We begin with a far-field study of light coupling and propagation in a MoS$_2$ waveguide, figure 1a. Figures 2a and 2b show optical and scanning electron microscopy (SEM) images of the waveguides fabricated with standard nanofabrication techniques atop a mechanically exfoliated MoS$_2$ flake on a SiO$_2$/Si substrate (oxide layer is 2 μm thick). In our far-field experiments, the waveguides are excited with input and output grating couplers optimized for the 950 nm – 1000 nm wavelength band (details of the design are available in Supplementary Sections II and VI). The length of fabricated waveguides is limited by the exfoliated flake size (typically < 40 μm × 40 μm area). Therefore, in order to increase the signal contrast between the output grating coupler and overall light scattering, we design our experiments for a cross polarization far-field microscopy (Supplementary Section II). For this purpose, a 90° bending with a radius of 3 µm is introduced. The fabricated waveguides are 85 nm thick and 320 nm wide (measured by atomic force microscopy (AFM) and SEM; a detailed description of device fabrication and characterization, including Raman spectroscopy and energy-dispersive X-ray spectroscopy (EDS), are available in Supplementary Section IV). The designed waveguides are single mode supporting only a single transverse electric (TE) exciton-dressed photonic mode (polarized along the MoS$_2$ crystallographic plane). The respective mode profile is shown in figure 2c. Large thickness of the SiO$_2$ layer and high index contrast with MoS$_2$ confine optical mode in the waveguide with minimal leakage to the substrate (figure 2c). The calculated overall mode area, $A_{eff} = \frac{(\iint_{-\infty}^{\infty}|E|^2\,dxdy)^2}{\iint_{-\infty}^{\infty}|E|^4\,dxdy}$, at 1 µm wavelength is ~0.053 $\mu m^2$ (~21% of a free space diffraction-limited spot, $A_0 = \lambda^2/4$). The fabricated MoS$_2$ waveguide size and respective mode area are significantly smaller than that of a corresponding optimum for a silicon waveguide (for comparison, in silicon minimum mode area of 0.072 $\mu m^2$ is attained for a 140 nm by 365 nm ridge waveguide, Supplementary Section III).

Upon illumination of the input coupler with the incident beam polarized along the grating ($\theta_{in} = 0^o$ as shown in figure 2f), TE exciton-dressed photonic mode [38, 39] is excited and guided in the waveguide (Supplementary Section I). Figure 2d shows a far-field confocal microscope image collected in a cross-polarized light centered around 915 nm (1.36 eV) excitation wavelength. Respective device transmission spectra are plotted in figure 2e. To understand the measurements, we perform detailed FDTD simulations (also plotted in figure 2e). The transmission spectra are a function of the spectral responses of the grating couplers, coupling efficiency, and optical extinction in the waveguide itself. Measurements and simulations show that no light is guided and transmitted below 900 nm, which we attribute to a strong light absorption above bulk bandgap (1.29 eV) [56] and close to A exciton resonance at 670 $nm$ (1.85 eV) [47] (Supplementary Sections II and VI), as well as to a high index contrast, which reduces the coupling efficiency (Supplementary Section II). Moving closer to the band edge, optical extinction in MoS$_2$ decreases, and grating coupler efficiency increases, making light transmission through the device possible. The observed peak in transmission spectra at ~980 nm is associated with the grating coupler efficiency optimum (Supplementary Section VI). Worth noting that due to strong mode confinement in the waveguide core, no notable light scattering is observed at the waveguide bend despite a small bending radius. In contrast, strong scattering is seen at the taper—waveguide interface, suggesting that a better-optimized coupler geometry is needed [57].

To further verify that the fabricated waveguide is single mode and only TE exciton-dressed photonic mode can be excited, we perform polarization-dependent spectroscopy as plotted in figure 2f. Specifically, we vary incident light polarization from $\theta_{in} = 0^o$ to $\theta_{in} = 90^o$ (output light is collected at $\theta_{out} = 0^o$ with respect to the output coupler) and observe that light transmission is reduced, manifesting that only TE waves are guided (observed small signal at $\theta_{in} = 90^o$ is attributed to a light scattered by the input port). The measurement agrees well with FDTD simulations (Supplementary figure S22). Supplementary Section VII further discusses the measurement of control devices with an input-only grating and expands on polarization-dependent waveguide excitation.

Next, to gain a better understanding of wave propagation within such deeply subwavelength MoS$_2$ devices, we perform near-field nano-optical imaging of a set of straight 15 $\mu m$, 20 $\mu m$ and 25 $\mu m$ long waveguides (fabricated waveguides are 80 nm thick and 300 nm wide, see figure 3b and Supplementary Section VIII). A schematic illustration of the setup is shown in figure 3a. In the experiment, we use a scattering-type scanning near-field optical microscope (s-SNOM) with a spatial resolution of ~25 nm (i.e., ~12 points across the waveguide) [58]. Near-field imaging and spectroscopy have emerged as a powerful tool to examine the dispersion of waves in a diverse set of materials [59-62]. Prior near-field studies of wave propagation in exfoliated flakes have revealed strong exciton-photon coupling and highly anisotropic wave propagation [63,64]. At the same time, to the best of our knowledge, the near-field study of nanostructured transition metal dichalcogenide devices has not been performed yet. In the case of $MoS_2$, such a measurement is challenged by a small device footprint and by a need for s-polarization for waveguide excitation (to excite the TE exciton-dressed photon mode). These factors contribute to a weak signal contrast necessitating the careful design of the experiment.

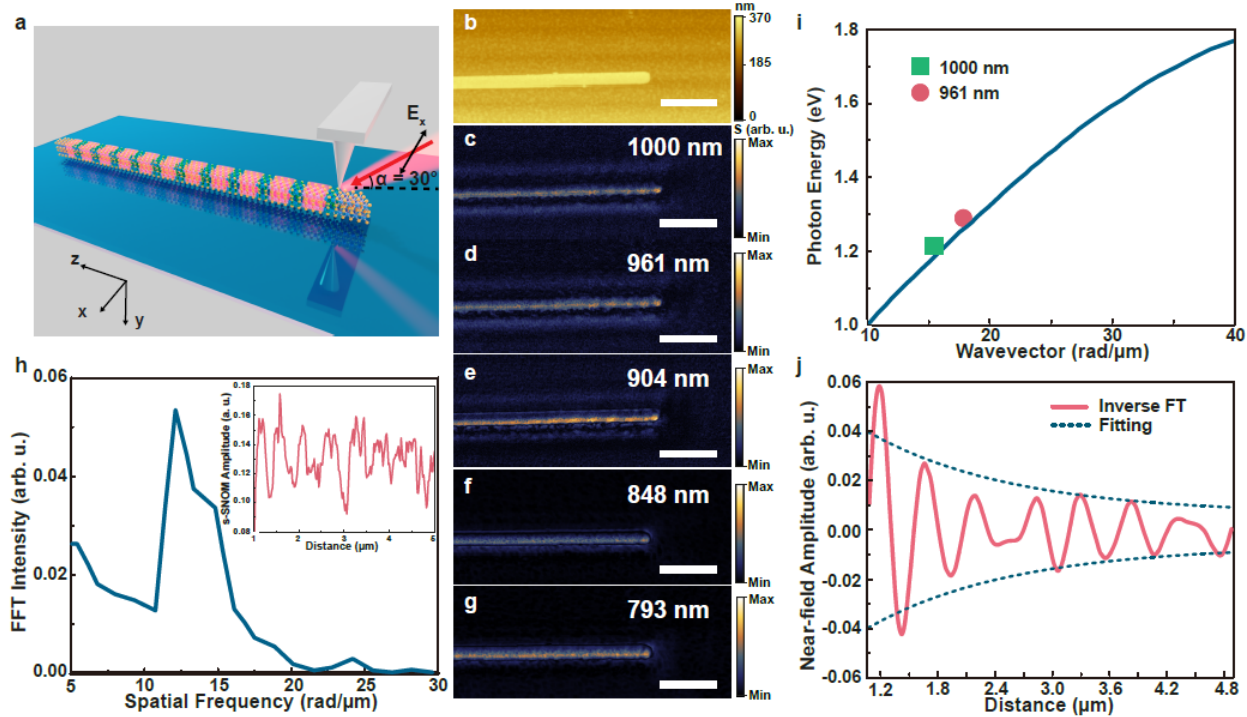

*Figure 3. Near-field study of deeply subwavelength $MoS_2$ waveguides. **(a)** Conceptual illustration of the setup. **(b)** AFM topographic map of the device. **(c-g)** s-SNOM images at wavelengths ranging from 793 nm to 1000 nm. **(h)** Fourier transform spectrum of interference fringe line profile extracted at 961 nm excitation wavelength (inset). **(i)** Predicted and extracted TE exciton-dressed photonic mode dispersion. **(j)** Inverse Fourier transform of the FT spectrum shown in panel **(g)** from 7 rad/μm to 20 rad/μm. A fitted exponentially decaying envelope is also shown.*

Figures 3c-3g show a set of s-SNOM images collected at various excitation wavelengths for 15 μm long waveguide under an *s*-polarized excitation [55] (that is, the incident electric field is perpendicular to the $MoS_2$ crystal axis and waveguide, figure 3a). Similar to the far field measurement (figure 2e), below 900 nm waves are not excited. Above 900 nm, we observe interference fringe patterns at both edges of the waveguide, signifying that guided TE exciton-dressed photonic mode is excited and guided. These interference fringe patterns are generated by a superposition between light scattered directly off the s-SNOM tip and excited mode [63] (see also Supplementary Section VIII). To extract guided wave dispersion and related insertion loss, we perform spatial Fourier transform on collected interference fringe profiles [39, 65]. Figure 3h

(inset) shows the s-SNOM amplitude of an interference fringe profile and a related spatial Fourier spectrum for a waveguide excited at 961 nm (1.29 eV). By tracing the peak of the Fourier spectrum, we extract the interference fringe period, $\rho$, which is related to the wavelength of the excited TE exciton-dressed photonic mode: $\lambda = \frac{\rho \lambda_0}{\rho \cos\alpha + \lambda_0}$ (see Supplementary Section VIII for details), where $\alpha = 30^o$ is the incidence angle (figure 3a) and $\lambda_0$ is the excitation wavelength [63]. In figure 3i we plot theoretically predicted wave dispersion and dispersion extracted from the near-field measurement, where an excellent agreement is revealed. To better visualize interference as a function of tip position we filter out short ($< 7 \, rad/\mu m$) and long ($>20 \, rad/\mu m$) spatial frequencies (figure 3j). Observed exponentially decaying envelope is then used to estimate the propagation length, $L_p$, of the excited TE mode (Supplementary Section VIII). We obtain $L_p \simeq 1.54 \, \mu m$ at 961 nm (1.29 eV) and $L_p \simeq 2.94 \, \mu m$ at 1000 nm (1.24 eV). Observed propagation lengths are significantly smaller than the theoretical prediction ($> 100 \, \mu m$) [45, 46], which we attribute to sidewall and surface roughness of fabricated devices [39]. We expect that as TMDC nanostructure fabrication processes are matured device performance will get close to a theoretical limit. Furthermore, in this study we focus on examining wave dispersion close to A exciton and across the bandgap, where optical extinction is naturally high. Operating at longer wavelengths ($\geq 1200 \, nm$) is expected to yield low loss functional deeply subwavelength devices.

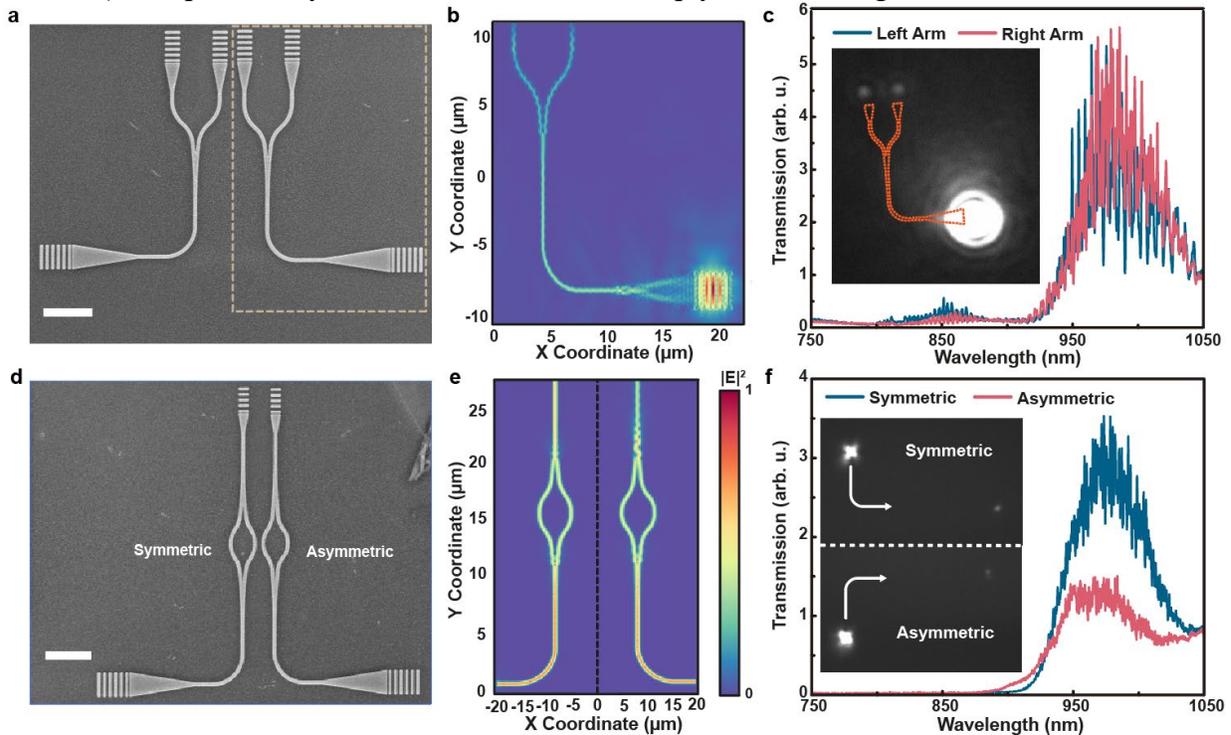

*Figure 4. Integrated exciton-driven MoS$_2$ nano-photonic devices. (a) SEM image of Y-junction beam splitters. Scale bar: 5 µm. (b) Simulated field intensity for a device shown in (a) at $\lambda = 950 \, nm$ excitation wavelength. (c) Beam splitter transmission spectra. Inset: Far field image of the Y-splitter collected at 915 nm center wavelength. (d) SEM image of symmetric and asymmetric Mach-Zehnder interferometers. Scale bar: 5 µm. (e) Simulated field intensity for interferometers shown in (d) at $\lambda = 950 \, nm$. (f) Transmission spectra for symmetric and asymmetric Mach-Zehnder interferometers, respectively. Inset: Far field image of Mach-Zehnder interferometers at 915 nm center wavelength.*

Next, to elucidate the capability of MoS$_2$ for deeply subwavelength integrated nano-photonics, we perform a far-field study of several passive photonic devices. Figure 4a shows an SEM of a fabricated 60 nm thick (i.e., $< \lambda/16$ at 1000 nm), 360 nm wide, and ~33 $\mu m$ long Y-junction beam splitter (see Supplementary Section IX for details). For such a small device cross-section only a fundamental TE exciton-dressed photonic mode can be excited. In figure 4b, we plot a corresponding profile of the wave excited at $\lambda = 950\ nm$ (figure 1b shows corresponding mode dispersion with and without exciton contribution). A respective far-field cross-polarized confocal image is shown in the inset of figure 4c. Two equally bright emission spots are observed at the output couplers (spaced 5 μm apart) of the device. figure 4c shows corresponding spectra measured at both output couplers. The analysis of overall power carried in the 900 nm to 1050 nm band (i.e., $\int_{900}^{1050} T_i(\lambda)d\lambda$, where $T_i(\lambda)$ is the transmission spectral power density through either the left or right arm of the beam splitters) yields a 47.15/52.85 splitting ratio between the two outputs, which indicates that a nearly 50/50 splitting of optical power is attained, as designed.

Based on the Y-splitter design, we further study a Mach Zehnder interferometer. The symmetric (with equal length of arms of ~10.41 μm) and asymmetric (shorter arm length is ~9.9 μm, i.e., ~500 nm arm length difference) interferometer structures with an overall device length of ~40 $\mu m$ are designed (figure 4d). Figure 4e shows respective simulations of field profiles at $\lambda = 950\ nm$. Constructive interference in the case of a symmetric structure is clearly seen, although the overall amplitude decreases due to optical loss at this wavelength. On the other hand, the phase difference between the arms of the asymmetric device is designed to yield destructive interference at the output. We note that due to an appreciable optical extinction at 950 nm, the two arms lead to a different insertion loss. As a result, a complete phase cancellation is not attained in the wavelength band studied here (750 nm – 1050 nm). Figure 4f presents a corresponding far-field study of fabricated symmetric and antisymmetric interferometers. Cross-polarization images (figure 4f, inset) suggest a stronger attenuation by an antisymmetric device, as expected. The difference between symmetric and antisymmetric configuration is further seen from measured transmission spectra (figure 4f). The observed asymmetry in transmission at ~980 nm excitation wavelength is more than two times and matches well with theoretical predictions (Supplementary Section IX).

To conclude, we demonstrated through simulations, far-field and near-field measurements that MoS$_2$, owing to its unique exciton-driven dispersion, is well suited for deeply subwavelength integrated nanophotonics. Functional devices that confine, control and guide light in devices with dimensions $<\lambda/10$ can be crafted. While in this work, we have focused on the 750nm –1050 nm wavelength band, the principles discussed here can be extended to a telecom band (>1200 nm) and other excitonic van der Waals materials [21] to create low loss [45-49] deeply subwavelength integrated photonic devices [30, 46]. Combined with the unique electronic and optical properties of 2D materials, exciton-driven subwavelength systems are of great promise to expand the nano-photonics applications palette [30].

**Acknowledgements**
A.R.D. acknowledges support of UCLA Society of Hellman Fellows, and partial support of NASA NIAC Program (grant # 80NSSC21K0954). A.M. acknowledges support by NSF (Grant: NSF-1845009 and NSF-ECCS-1708579). S.D. acknowledges support from NSF under grant DMR-2005194, OIA-2033454, and ACS PRF fund 66229-DNI6.

**Contributions**
H.L. and A.R.D designed the structures. H.L. fabricated and characterized samples, developed theoretical background and performed simulations. H.T.T performed SEM and XRD measurements. Arn. M., A.M., H.L., and A.R.D. planned far-field measurement. A.M. supervised far-field experiment. Arn.M., D.S. and J.F. performed far-field microscopy, spectroscopy and data collection. S.D., J.S., H.L., and A. R. D. planned near-field experiment. S.D. supervised near-field experiment. J.S. performed near-field imaging and data processing. A.R.D initiated and coordinated research project. All the authors were involved in the discussion of results and the final manuscript editing.

# Supplementary Information

## I. High refractive index and principles of deeply subwavelength guiding.

As detailed in the main text, at present dielectrics and metals constitute the two major avenues for creating subwavelength photonic devices. Below we provide a brief discussion of light confinement in conventional dielectrics, metals, and excitonic MoS₂.

### I.1. Dielectrics.

In dielectric structures device size and light confinement are governed by the material refractive index, $n$. Optimal device footprint can be approximately assessed as $\frac{\lambda}{2n}$, corresponding to a diffraction limited spot in the medium (here $\lambda$ is the wavelength). Of a particular interest are optical devices that can operate at photon energies below the material bandgap ($\lambda > \lambda_g$), where optical extinction coefficient, $k$, is small (i.e., $n \gg k$ and $k \to 0$). Survey of key presently used photonic materials (figure S1) shows that traditional optical materials yield at most $\sim\lambda/7$ confinement in the near-IR band (e.g., for Si with a refractive index of ~3.5). Deeply subwavelength optics discussed in the main text (i.e., with device sizes $<\frac{\lambda}{10}$) is challenging to attain with conventional covalent dielectric materials.

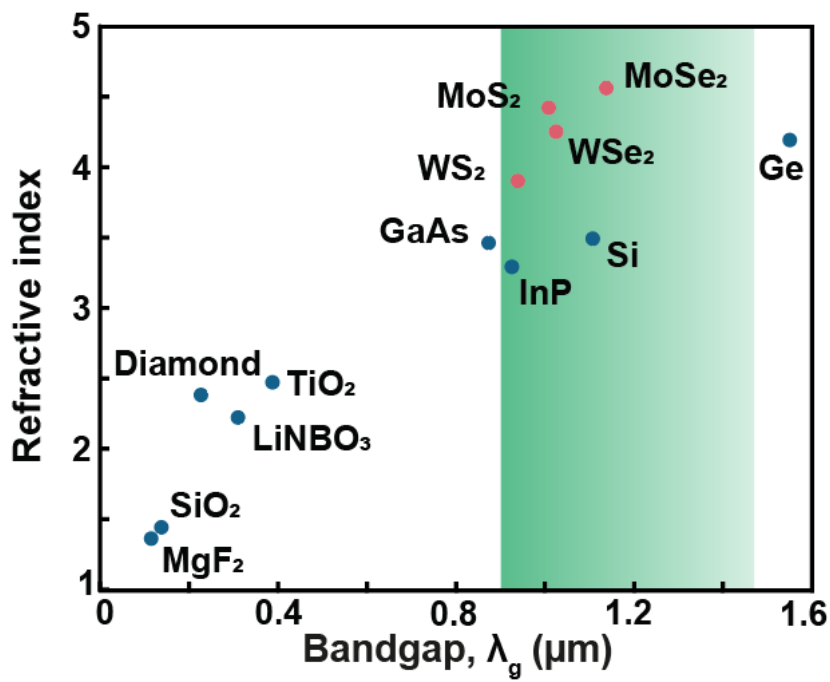

*Figure S1. Refractive index as a function of material bandgap. Shaded area denotes the near-IR wavelength range of interest. Bulk transition metal dichalcogenides are highlighted with red markers.*

Understanding the connection between refractive index and material electronic structure has been extensively examined in a number of prior works [1-5]. While the refractive index, $n$, varies greatly across different materials, several generic conclusions can be made from fundamental Kramers-Kronig relations. In particular, it was shown [2, 4, 5] that below bandgap refractive index can be approximated as:

$$n(\lambda) \simeq 1 + \int_0^{\lambda_g} K(\xi)d\xi. \qquad (1)$$

where $\lambda \gg \lambda_g$ [4,5] is assumed, $K(\lambda) = \frac{4\pi k(\lambda)}{\lambda}$ is the material absorption coefficient, $k(\lambda)$ is the extinction coefficient, $\lambda_g$ is the wavelength of material optical bandgap.

Evidently, below-bandgap refractive index depends on the material absorbance, $K(\lambda)$, and the optical bandgap, $\lambda_g$. Specifically, materials with a larger bandgap (i.e., smaller $\lambda_g$) typically possess a smaller refractive index, whereas smaller bandgap materials, such as germanium (Ge) exhibit high refractive index values [2, 4]. In figure S1 we plot average sub-bandgap refractive index values as a function of material bandgap for several typical materials used in photonics (data is based on tabulated values). While narrow gap materials (such as Ge) yield a high refractive index, they are not suitable for nano-optics due to a high absorption in the near-IR range.

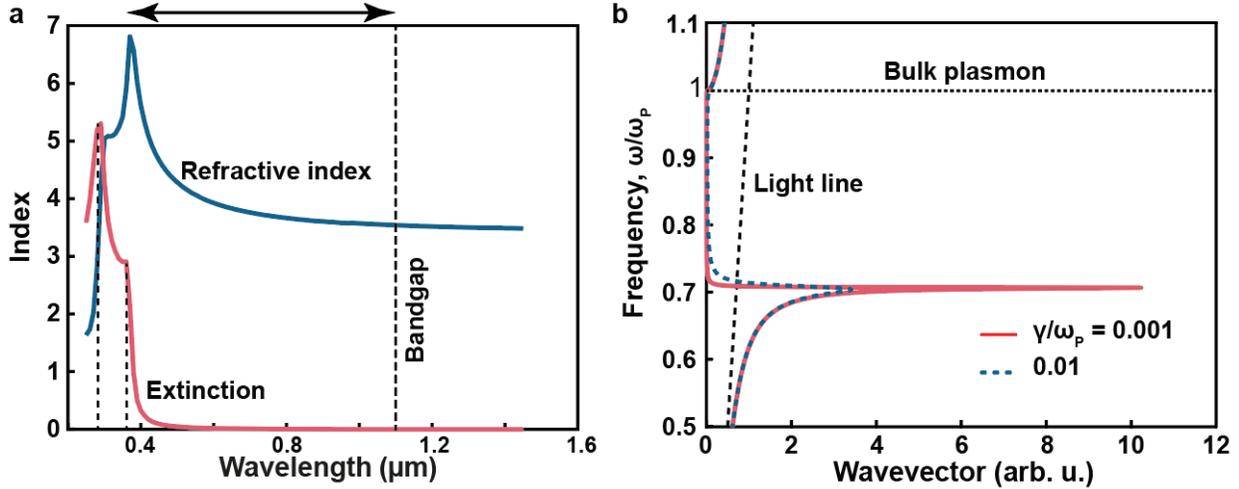

*Figure S2. Optical responses of dielectrics and metals. a, Dispersion of real and imaginary (extinction) parts of refractive index of Si. Below-bandgap refractive index (at $\lambda > 1100\ nm$) is dominated by strong UV interband transition resonances. b, Surface plasmon wave dispersion for two different values of damping coefficient $\gamma$. Strong coupling between bulk plasmons and free-space photons is seen.*

The limitation on the refractive index value can be understood from a study of materials optical extinction, $k(\lambda)$. Detailed analysis shows that dominant contribution to optical extinction is due to strong interband transition resonances [5, 6] associated with a high joint density of states over a large Brillouin volume, and are observed well above the optical band edge. For example, in silicon (Si) these transitions occur < 400nm, that is well above the bandgap at 1100 nm (see figure S2a). In gallium arsenide (GaAs), despite it being a direct band semiconductor, strong transitions are observed only at $\lambda < 500\ nm$ (i.e., > 400 nm above the optical absorption edge) [6]. A large "gap" between the onset of optical band edge ($\lambda_g$) and extinction resonances [3] results in a limited value of the below-bandgap refractive index. This simplistic analysis further suggests that materials with strong resonances close to the band edge are expected to have a higher sub-bandgap refractive index. As we show in Sec. I.3., excitons in transition metal dichalcogenides help mediating high refractive index [1].

Overall, a brief survey of conventional covalent materials shows that achieving deeply subwavelength optical devices is challenging with conventional dielectrics. We would like to also refer to recent papers [1, 7], where other limitations are discussed. At the same time excitonic transition metal dichalcogenides [8], such as MoS2, demonstrate a much higher index of refraction [9], figure S1, which allows creating devices $< \lambda/10$.

### I.2. Metals and surface plasmon-polaritons.

Metals provide a conceptually different approach for deeply subwavelength photonics. Optical response of metals can be modelled by a free electron Drude plasma model (assuming that the wavelengths of interest are outside of interband transition range) [10, 11]. In this case the permittivity is given as $\varepsilon = 1 - \frac{\omega_p^2}{\omega(\omega+i\gamma)}$ (here for simplicity we assume unity background permittivity, $\varepsilon_\infty = 1$), $\omega_p$ is the plasma frequency and $\gamma$ is the collision frequency responsible for damping. Free electron plasma has a natural resonance at $\omega_p$, where $\varepsilon \to 0$, associated with the excitation of longitudinal charge density oscillations – bulk plasmons. At a metal surface such longitudinal oscillations can couple to electromagnetic waves with a formation of surface plasmon-polaritons – hybrid states between optical photons and charge density oscillation quanta [10, 11]. Importantly, the strong coupling between photons and charge density oscillations leads to significantly altered dispersion of guided waves enabling high wavenumber modes (large effective guide index) near the surface plasmon resonance frequency [12-16]. For the sake of completeness of this discussion in figure 2Sb we plot a dispersion of surface plasmon modes for two different values of damping coefficient, $\gamma$. It is the excitation of these modes with large wavenumbers, which possess effective guide indices >10 times higher than that of free space modes, is used to confine and guide light deep beyond the diffraction limit [17, 18]. However, unlike dielectrics discussed above, in metals excited states of free electron gas are inherently associated with enhanced optical absorption [10, 19, 20]. Furthermore, the stronger is confinement (i.e., the closer one gets to a surface plasmon resonance) the higher is the associated loss [16]. Inevitable loss of plasmonic devices naturally limits the scope of potential applications.

The aforementioned analysis of dielectrics and metals shows that it is desirable to combine the advantages of both worlds. Specifically, materials with a bandgap such as semiconductors are low loss below the optical absorption edge, whereas achieving strong resonances, and potentially strong coupling, close to the band edge may enable high refractive indices. Below we show that excitonic materials, such as MoS2 studied in this work, are ideally suited for this purpose.

### I.3. Refractive index of excitonic semiconductors and exciton – photon coupling. Case of MoS2.

Transition metal dichalcogenides possess room temperature exciton resonances that are manifested as strong absorption peaks in optical extinction spectra, see figure S3 [9]. Of a particular interest for us is MoS2, studied in the main text. To understand the role and contribution of excitons to the observed wave dispersion, we model previously measured [21] optical constants of MoS2 by making use of Tauc-Lorentz oscillator model of permittivity. Specifically, we model the material permittivity as a sum of 6 Tauc-Lorentz oscillators [21, 22], corresponding to physically observed absorption resonances:

$$\varepsilon_2(E) = \sum_i \frac{A_i E_{0i} C_i (E - E_{gi})^2}{(E^2 - E_{0i}^2)^2 + C_i^2 E^2} \frac{1}{E} H(E_{gi})$$

$$\varepsilon_1(E) = \varepsilon_b + \frac{2}{\pi}\mathcal{P}\int_{E_g}^{\infty}\frac{\xi\varepsilon_2(\xi)}{\xi^2 - E^2}d\xi$$

where $\varepsilon_2$ corresponds to imaginary part of permittivity, $A_i$ is the strength of *i*-th oscillator, $E_{oi}$ is the transition energy of the oscillator, $C_i$ is the broadening term of the *i*-th oscillator, $E_{gi}$ is the optical band gap of associated with a given *i*-th oscillator, $H(x)$ is the Heaviside step function, $\varepsilon_1$ is the real part of permittivity obtained via Kramers-Kronig relation (Cauchy principal value of the integral is assumed), and $E = \hbar\omega$ is the photon energy ($\omega$ is the angular frequency).

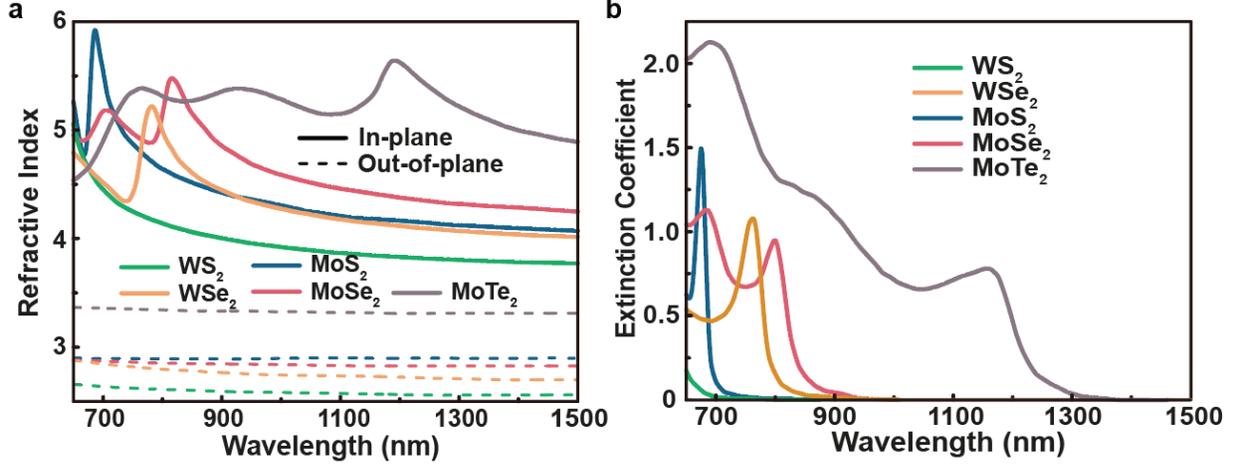

*Figure S3. Optical constants of bulk TMDCs. Refractive index along both in-plane and out-of-plane orientations, **a**, and in-plane extinction coefficient, **b**, for bulk WS₂, WSe₂, MoS₂, MoSe₂ and MoTe₂ after Refs. [9].*

We then fit the oscillator model to the experimental data [21] with the use of genetic algorithm optimization [23]. Specifically, we search for a minimum of the following objective function:

$$F = \left|\sum_{E_q}\left(\varepsilon_{1q} - \varepsilon_1(E_q)\right)\right| + \left|\sum_{E_q}\left(\varepsilon_{2q} - \varepsilon_2(E)\right)\right|,$$

where $E_q$ corresponds to photon energies of measured data points, $\varepsilon_{1q}$ and $\varepsilon_{2q}$ are measured real and imaginary refractive indices of MoS₂, respectively. Figure S4a shows a comparison between a fitted model and measured data [21].

Once the permittivity model is established, we determine contribution of MoS₂ A and B excitons to the refractive index. In figure S4b we plot refractive index as a function of a wavelength with and without contribution of A and B excitons. The extracted difference in refractive index is $\Delta n \simeq 0.3$ at $\lambda = 1\ \mu m$.

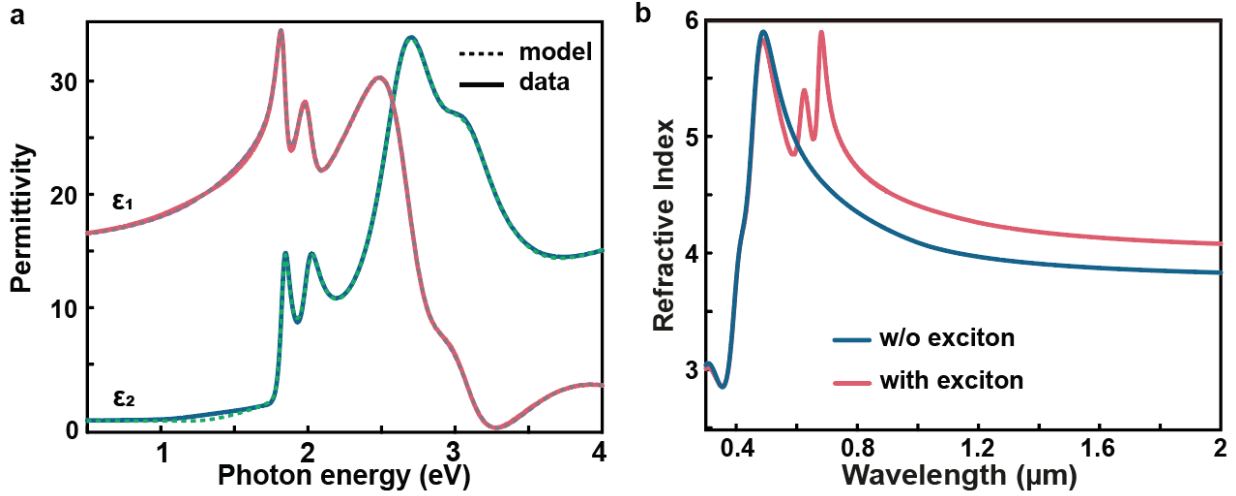

*Figure S4. Role of excitons in MoS$_2$ refractive index.* **a**, *Comparison of oscillator model and measured data [21].* **b**, *Refractive index with and without A and B exciton contribution.*

Moving forward we study exciton – photon coupling and examine related wave dispersion. We note that exciton-photon coupling was studied in a range of prior works [24-32]. Strong coupling and formation of exciton-polaritons were examined. Here, we look at the nature of such coupling in bulk MoS$_2$ and show that exciton-photon coupling is naturally manifested in MoS$_2$.

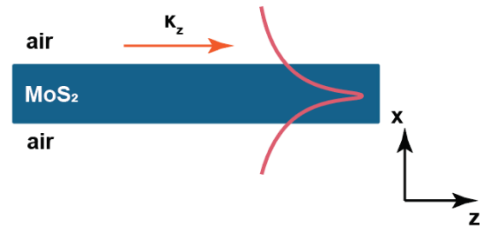

*Fig. S5. MoS$_2$ slab waveguide.*

To elucidate exciton-photon coupling in bulk MoS$_2$, we consider a case of a simple 70 nm thick MoS$_2$ slab waveguide in air, schematically shown in figure S5. Based on the Tauc-Lorentz oscillator model [21, 22], we express the MoS$_2$ complex permittivity as $\varepsilon(E) = \varepsilon_{wo}(E) + \Delta\varepsilon(E)$, where $\varepsilon_{wo}(E)$ is the permittivity without exciton contribution, $E$ is the photon energy. Without excitons the slab would guide optical modes with wave dispersion governed by $\varepsilon_{wo}$, whereas in the presence of excitons dispersion is driven by $\varepsilon$. In both cases, the guided wave dispersion is sought as an eigen-value problem at a given photon energy $E$ yielding a complex wavevector $\tilde{\kappa}_{guide}$ (respective dispersions are plotted in figure 1b of the manuscript and in figure S6 below).

However, to examine intrinsic coupling between A, B excitons and guided photon modes, we study excitation of the slab. In this case we assume that a plane wave $\propto e^{i\sqrt{\left(\frac{E}{\hbar c}\right)^2 \varepsilon_{air} - \kappa_z^2}}$ is incident on a slab and explore extinction in the slab as a function of the incident wave wavevector, $\kappa_z$, and photon energy, $E$. For $\kappa_z < \frac{E}{\hbar c}\sqrt{\varepsilon_{air}}$ this study would correspond to examining absorption in the slab as a function of the angle of incidence of the plane wave. Such a study has been performed experimentally in a number of prior works [24, 33]. Here, however, we extend analysis to $\kappa_z > \frac{E}{\hbar c}\sqrt{\varepsilon_{air}}$, which corresponds to a slab excitation with evanescent waves. Such excitation is performed, for example, in the study of plasmon-polaritons and phonon-polaritons and can be achieved with Otto configuration or near-field tip excitation [1, 34, 35]. Importantly, extension to evanescent wave excitation enables to examine excited modes in the entire $\kappa_z - E$ plane and obtain momentum energy loss spectra. In figure S6a we plot optical extinction (calculated as

$\frac{E}{\hbar} Im(\varepsilon(E)) \int_o^h |\mathcal{E}|^2 dx$, where $\mathcal{E}$ is the excited electric field, in our case polarized along the y-axis, $h$ is the thickness of the slab).

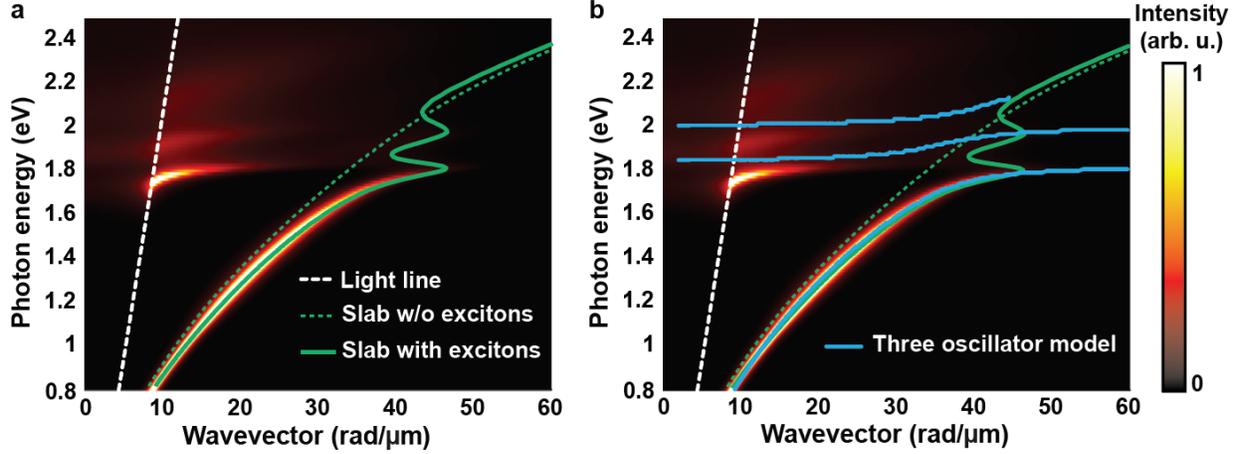

*Figure S6. MoS$_2$ slab excitation.* The color maps show wave extinction in the slab. In addition, panel **a** shows dispersion of the slab waveguide with and without contribution of A and B excitons; panel **b** shows resonance spectra based on a three-oscillator model.

The observed spectra clearly show three excited branches: the guided wave, and A and B excitons. Notably, a strong dispersion (i.e., dependence on the wavevector $\kappa_z$) of A and B exciton branches is seen, which signifies that excitons are coupled with photon modes in a slab of bulk MoS$_2$. Indeed, exciton radius is much smaller than the photon wavelength, as a result, without intrinsic exciton-photon coupling, dependence with the $\kappa_z$ would not have been observed.

To further examine the nature of the coupling between excitons and photon, we develop a simplistic three oscillator model:

$$\ddot{x}_{ph} + \gamma_{ph}(\kappa_z)\dot{x}_{ph} + E_{ph}(\kappa_z)x_{ph} - \eta_{ph,A}x_2 - \eta_{ph,B}x_3 = F_1 e^{-i\frac{E}{\hbar}t}$$

$$\ddot{x}_A + \gamma_A \dot{x}_A + E_A x_A - \eta_{A,ph} x_{ph} = F_2 e^{-i\frac{E}{\hbar}t}$$

$$\ddot{x}_B + \gamma_B \dot{x}_B + E_B x_B - \eta_{B,ph} x_{ph} = F_3 e^{-i\frac{E}{\hbar}t}$$

here oscillator $x_1$ corresponds to an amplitude of an exciton-free guided photon mode (i.e., with dispersion obtained from $\varepsilon_{wo}(E)$), are the decay rate and energy of the photon mode, both $\gamma_{ph}(\kappa_z)$ and $E_{ph}(\kappa_z)$ depend on the wavevector $\kappa_z$ and are found from the exciton-free slab dispersion (i.e., $\tilde{\kappa}_{guide}(E)$), oscillators $x_A$ and $x_B$ correspond to A and B excitons, $\gamma_A$, $\gamma_B$, $E_A$ and $E_B$ are their intrinsic decay rates and eigen-energies, respectively, $\eta_{ph,A} = \eta_{A,ph}$ and $\eta_{ph,B} = \eta_{B,ph}$ are respective exciton-photon coupling strengths, and, lastly, $F_i e^{-i\frac{E}{\hbar}t}$ are external driving fields.

In our modelling we assume that $F_2 = F_3 = 0$. Exciton decay rates, $\gamma_{A,B}$ can be estimated from experimental measurements [36, 37], here we assume ~1 *ps* lifetime (~1 *meV*). The photon decay rate is estimated as $\gamma_{ph}(\kappa_z) \simeq E_{ph}(\kappa_z)\frac{Im(\tilde{\kappa}_{guide})}{Re(\tilde{\kappa}_{guide})}$ based on the propagation length of guided optical modes (where $\frac{Im(\tilde{\kappa}_{guide})}{Re(\tilde{\kappa}_{guide})}$ factor corresponds to a quality factor of the exciton-free guided wave). In this case, the only free modelling parameter is the coupling strengths $\eta$. By varying $\eta_{ph,A}$ and $\eta_{B,ph}$ we find that three oscillator model predicts well the observed slab extinction when both coupling strengths are in the range 0.55 – 0.75 eV. The resonances of the three oscillator model are plotted in figure S6b, here $\eta_{A,ph} = 0.65\ eV$ and $\eta_{A,ph} = 0.6\ eV$ are assumed. A good match between a simple model and a full-wave simulation is seen. Based on these estimates we obtain intrinsic photon loss of $\gamma_{ph} \simeq 0.2\ eV$ at $E = 2\ eV$, $\gamma_A \simeq \gamma_B \sim 1\ meV$. Comparing with the strong coupling condition $2\eta > (\gamma_{ph} + \gamma_{exciton})/2$ we get that the studied system is in the strong coupling regime [38] (even in the case a stricter condition for estimating photon decay rate is used $\gamma_{ph}(\kappa_z) = E_{ph}(\kappa_z)\frac{2\pi\ Im(\tilde{\kappa}_{guide})}{Re(\tilde{\kappa}_{guide})}$). As a result, the guided waves can be identified as a lower exciton-polariton branch. We stress that this strong coupling is inherent to MoS2 and is only loosely related to structure. That is, excitons and photons are self-hybridized in the bulk of the MoS2 primarily due to the strong exciton dipole moment and large exciton density [25, 38, 39]. Optical design can further enhance and modify such a coupling. This manifestation of strong exciton-photon coupling has been also seen and reported in various other studies. Interestingly, the strong coupling exists even in the case when optical loss is approaching the photon energy, i.e., in a strongly damped optical system.

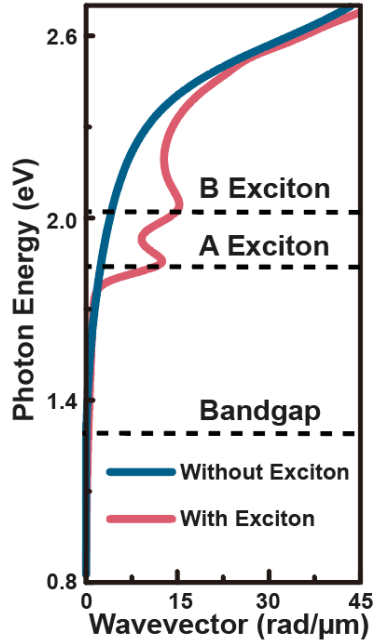

*Figure S7. Dispersion of imaginary part of the wavevector for MoS2 waveguide of 60 nm thickness and 360 nm width with and without exciton contribution.*

### I.4. Dispersion of the 60 nm MoS2 waveguide

In figure 1b of the main text we provide a dispersion of a 60 nm by 360 nm MoS2 waveguide on top of a SiO2/Si substrate. For completeness of our discussion, in figure S7 we plot also the dispersion of the imaginary part of the wavevector of the guided wave (i.e., $Im(\tilde{\kappa})$). The imaginary part of the wavevector corresponds to waveguide insertion loss. For example, one can estimate the propagation length as $L = \frac{1}{2Im(\tilde{\kappa})}$.

As seen in figure S7, in both cases of with and without exciton contribution, optical losses below the bandgap become small. We expect that devices with photon energies below 1eV (>1200 nm) will have a very small insertion loss. In our work we were interested in a study of wave propagation near A exciton and bandgap, as a result loss in our case is higher.

## II. MoS$_2$ waveguide and grating coupler design

As shown in figure S8a, the designed structure consists of two identical grating couplers for coupling light into and out of the waveguides. The period of the grating is 600 nm and the duty cycle is 0.267 (i.e., each bar of the grating is 160 nm in width). The width of the entire grating is designed to be 3 µm in order to cover the entire illumination spot of the incident beam. Connecting the grating coupler and the actual ridge waveguide are two tapers. We design them to be 7 µm in length to minimize light scattering at the taper – waveguide intersection point. Since we aim at working with device of deeply subwavelength scale, i.e., thickness < λ/10, the thickness of the flakes we used for fabrication ranges from 60 nm to 100 nm. To make sure only fundamental TE waveguide mode is excited and sustained, we set the width of such deeply subwavelength waveguide to be at 300 nm. In order to perform cross polarization far field measurement, we also introduce a 90° bend of a 3 µm radius. The overall length of the waveguide section, including the bend, is ~13.7 µm. FDTD simulations show that without taking into account material absorption, the loss of such bending in the waveguide is negligible, e.g., << 1%.

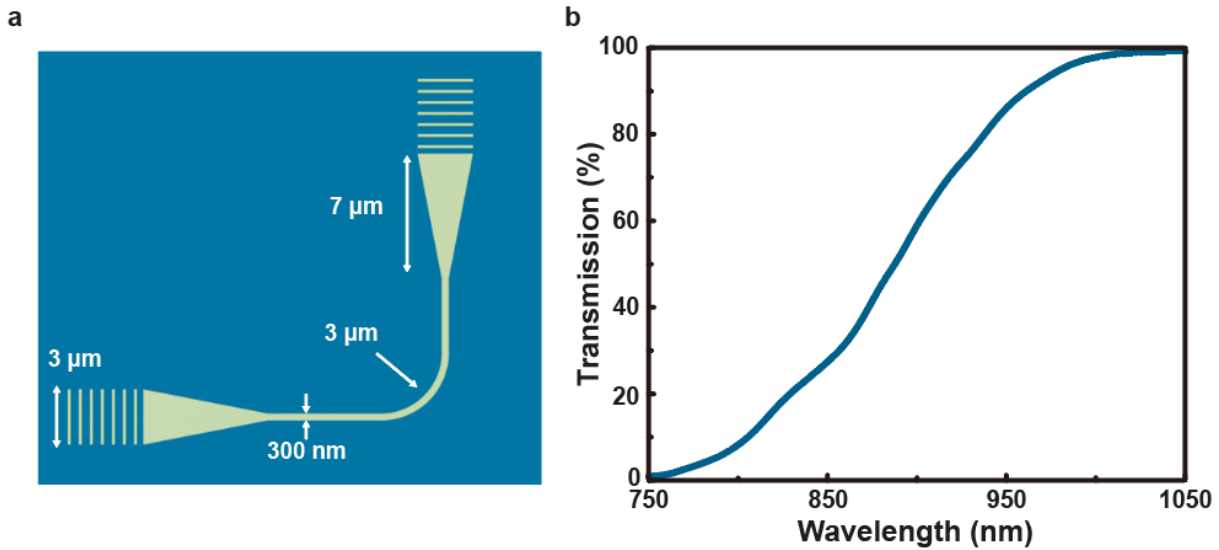

*Figure S8. **a**, Schematic illustration of a designed waveguide structure. **b**, Calculated transmission of the waveguide section shown in **a** (i.e., transmission from a taper tip of input coupler to the taper tip of the output coupler).*

Figure S8b shows calculated transmission of the waveguide section only. Close to A exciton resonance at ~680 nm optical extinction is large and transmission through the waveguide is nearly zero even though a relatively short 13.7 µm waveguide. Close to the band edge material loss becomes smaller and transmission through the waveguide grows reaching ~100 % for >1.2 eV.

In our experiments in addition to intrinsic material extinction loss, device performance is limited by imperfections of fabrication resulting in additional guided mode scattering and absorption on surface defects. It is reasonable to expect that as material growth is matured and better fabrication methods are developed, high quality sub-bandgap MoS$_2$ devices can be created.

Next we study input and output coupling efficiency of the grating couplers. First, we assess the input grating coupler efficiency. We illuminate the input grating with a 3 μm spot size Gaussian beam at a normal incidence (mimicking conditions of the experiment). We then examine the fraction of incident power coupled into the waveguide. Figure S9a demonstrates related input grating coupler efficiency. The efficiency of the input grating peaks at ~950 nm and is ~1.5 %. Low input grating coupler efficiency can be attributed to a significant mismatch between the device size ($\sim 60\ nm \times 300\ nm$) and the incident beam size, as well as a sharp refractive index contrast between MoS$_2$ and air. Optimization of grating performance is not the objective of this work, however we expect that with the use of advanced inverse design methods a more efficient grating design can be obtained.

Similarly, we assess the performance of the output grating. Specifically, we study power conversion from a guided wave to free space modes (here again a monitor atop of the output grating is placed). Figure S9a shows a related calculation. The output grating coupler efficiency is higher and also peaks at ~950 nm (~20 % at 950 nm).

Due to a strong material index dispersion, the coupling efficiency varies a lot, which also manifests in the output emission angle. In figure S9b we also plot the far field projection of the emission profile. At wavelength = 950 nm, the emission from the output coupler is almost vertically upward. As shown in figure S9b, at 900 nm wavelength, the emission angle is about 15°, and at 1050 nm, the angle is about -18°.

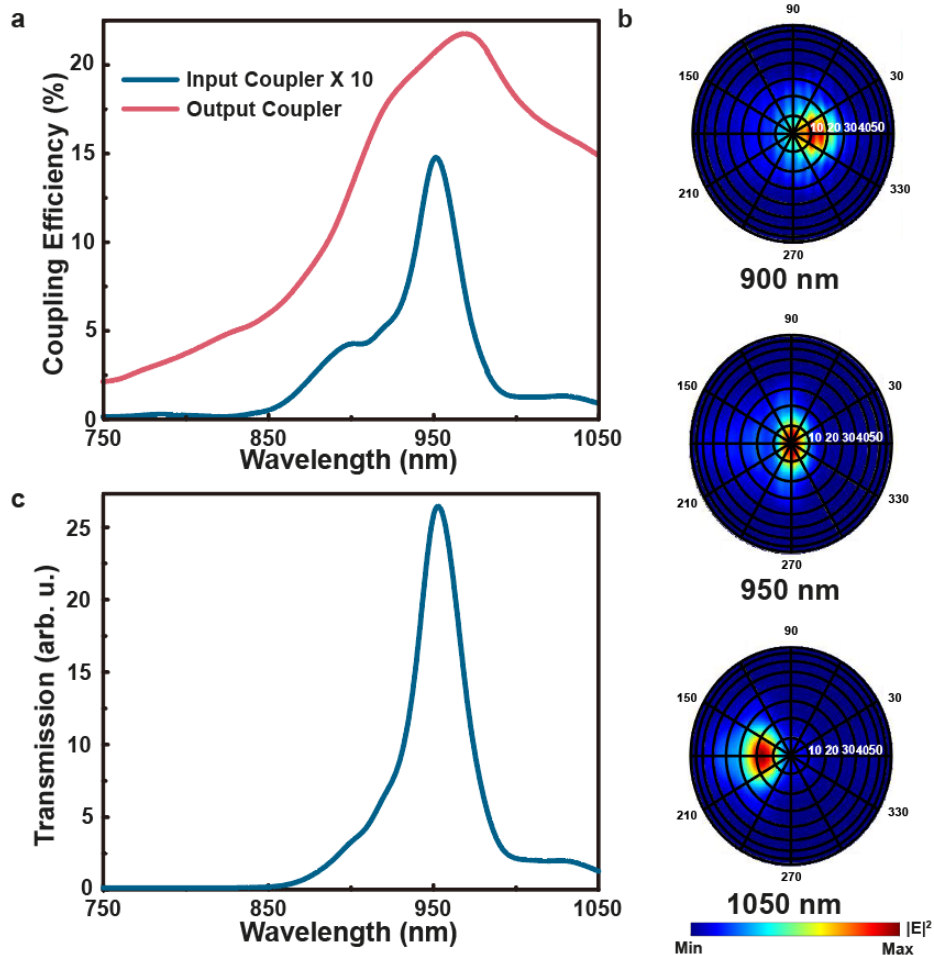

*Figure S9: Study of the waveguide and grating coupler performance. a, Coupling efficiencies*

of input and output grating couplers. *b*, Far field radiation patterns of the output grating coupler at three different wavelengths. *c*, Calculated transmission of the entire device shown in figure S8a.

The overall efficiency of the device shown figure S8a is then combination of three factors: $\xi_{total} = \xi_{input}\xi_{waveguide}\xi_{output}$. Figure S9c plots the calculated corresponding overall efficiency. The overall theoretical maximum of the designed device is ~0.2% at 950 nm. However, we note that in our case imperfections during fabrication process result in a slightly different fabricated geometry with modified performance parameters. Furthermore, imperfections of fabrication contribute to additional optical absorption. To account for fabrication imperfections on waveguide performance the SEM imaging is used to estimate dimensions of the fabricated device, which are then used to re-calculate coupler and waveguide performance (see Sec. VI and figure S15).

### III. Optical mode area and confinement factor in MoS$_2$ waveguide

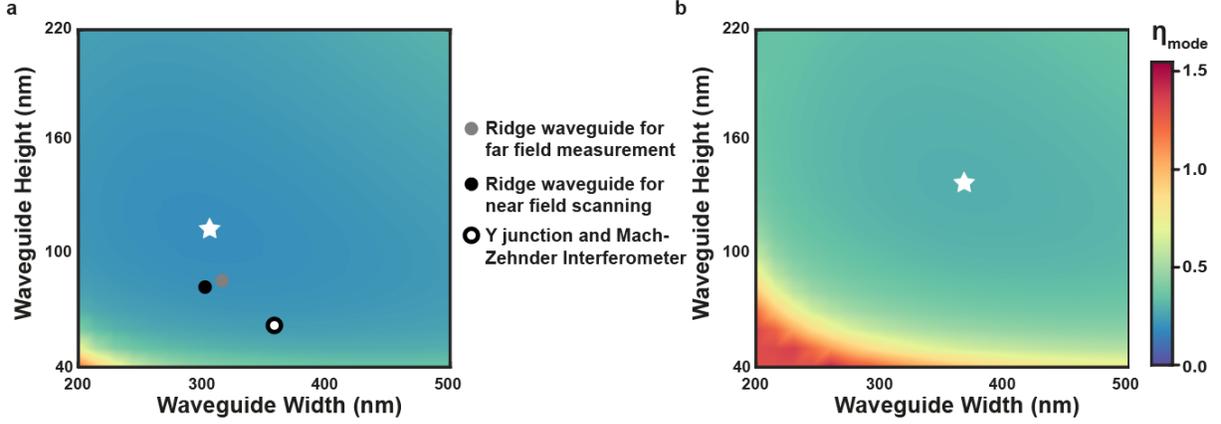

***Figure S10. Waveguide mode area study***. *Normalized mode area for fundamental TE mode in a MoS$_2$ waveguide (panel **a**) and in a Si waveguide (panel **b**) on SiO$_2$/Si substrate as a function of waveguide width and thickness. Here $\lambda = 1$ μm. The stars in both figures correspond to the optimal (smallest) mode area achieved in MoS$_2$ and Si devices, respectively. The markers in panel **a** correspond to geometric dimensions of the three different fabricated devices examined in this work.*

Since the fundamental TE mode is cut-off free it is important to investigate mode confinement to ensure optimal operation. Indeed, waves can be excited and guided even in a 1 atom thick waveguide [39]. However, in this case optical mode is highly delocalized. It is desirable to find an optimum between mode confinement and device footprint. For this purpose we study mode area in our waveguides as a function of device dimensions. We introduce effective mode area of the guided wave as:

$$A_{eff} = \frac{(\iint_{-\infty}^{\infty}|\boldsymbol{E}|^2\ dxdy)^2}{\iint_{-\infty}^{\infty}|\boldsymbol{E}|^4\ dxdy}$$

and define normalized mode area as $\eta_{mode} = \frac{A_{eff}}{A_0}$, where $A_0 = \frac{\lambda^2}{4}$ represents the approximated mode area of a diffraction limited spot in free space.

In figure S10a we plot normalized mode area for a fundamental TE mode at $\lambda = 1\ \mu m$ in MoS$_2$ waveguide as a function of waveguide dimensions. For comparison, we also plot $\eta_{mode}$ for fundamental TE guided mode in a Si waveguide, figure S10b. Higher refractive index of MoS$_2$ leads to stringer confinement of optical field and to smaller footprint devices, when compared with Si waveguide. With smaller and thinner devices, MoS$_2$ can more efficiently confine light at smaller effective area. The smallest mode area that can be achieved in a Si waveguide is > 50% larger than that in a MoS$_2$ waveguide (0.288 vs. 0.191, respectively). At the same time, corresponding cross-section for MoS$_2$ is 50% smaller than that of Si device (110 nm x 305 nm for MoS$_2$ vs 140 nm x 365 nm for Si; here dimensions are for optimal mode area in either of the materials).

We also mark the dimensions and the corresponding normalized mode area of three different fabricated MoS$_2$ devices: ridge waveguide for far-field measurement, ridge waveguide for near-field imaging, and Y-splitters and Mach-Zehnder interferometers (figure S10a). While fabricated devices are not globally optimal, corresponding normalized mode area for these three devices (~0.2 for waveguides and ~0.24 for Y-splitters and interferometers) is still smaller than the optimal value achieved for a Si waveguide (~0.288).

## IV. Fabrication process flow of MoS$_2$ integrated photonic devices on Si/SiO$_2$ substrates

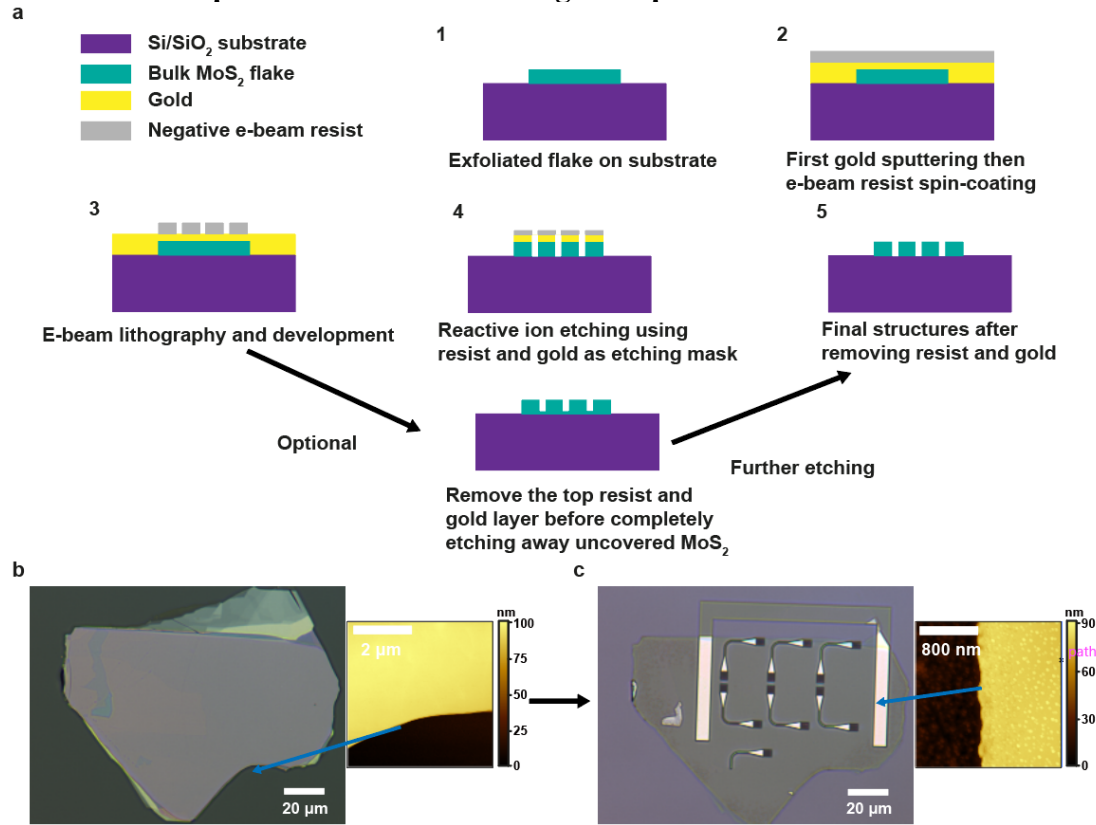

***Figure S11. MoS$_2$ integrated device fabrication procedure. a,*** *Schematic illustration of the fabrication process flow of MoS$_2$ waveguiding devices demonstrated in this work.* ***b,*** *Microscope optical image and AFM scan of as-exfoliated MoS$_2$ flake on Si/SiO$_2$ substrate.* ***c,*** *Microscope optical image and AFM scanning of the after-fabricated flake. The AFM scans in both **a** and **b** are performed in the region indicated by the arrows.*

The device fabrication process flow is shown in the figure S11a and follows prior reports [28, 40]. We begin by exfoliating MoS$_2$ flakes atop a SiO$_2$/Si substrate (2 μm wet thermal oxide; purchased from University Wafer). The flakes are exfoliated from bulk MoS$_2$ crystals (SPI

Supplies) with a standard scotch tape method. The topography and thickness of exfoliated flakes are then examined with optical microscopy (thickness of desirable flakes is predetermined based on color) and AFM (Bruker Dimension FastScan with ScanAsyst). Figure S11b shows an optical micrograph of an as-exfoliated flake and related AFM topographic profile (this flake is used for fabrication of devices measured in figure 2 of the main text).

After a large area flake of a desired thickness is determined (see figure S11b), we deposit a thin layer of gold about 20 nm (serves as a charge dissipation layer, an adhesion layer, and an etching mask) with sputter coating (Denton Vacuum Desk V Sputter). Next we deposit a thin layer of a negative e-beam resist, MaN 2403, with spin-coating technique. 300 nm thick resist is deposited with 3000 rpm for 30 seconds. Then the chip is baked at 95 °C for 1 minute. Then the flake is patterned by e-beam lithography (Raith EBPG 5000+ES) with 100 keV beam energy. After that, the written flake is developed with MF-319 solution followed by 90 seconds rinse in DI water. The resist exposed by e-beam would remain on the flake after development, serving as an etching mask as well. Next chlorine reactive ion etching (RIE) (PlasmaTherm SLR 770 ICP) is used to remove gold (recipe: 24 sccm Ar, 5 sccm $Cl_2$, RF power of 100 W and ICP RF power of 250 W, etching time 15s) and fluorine RIE (Oxford PlasmaLab 80+) is used to remove $MoS_2$ (recipe: 40 sccm $CHF_3$, 4 sccm $O_2$, forward power of 55 W and pressure at 40 mtorr). The estimated etching rate of $MoS_2$ is about 9 nm/second. After the exposed $MoS_2$ has been etched away, the remaining resist can be removed by immersion in hot acetone or oxygen plasma, and the gold layer can be washed away with Gold Etchant TFA.

One thing to notice here is that, during the fabrication, the final thickness of $MoS_2$ devices can be further tuned (the optional step shown in figure S11a). Instead of completely removing the exposed $MoS_2$, a thin layer of $MoS_2$ can be retained before washing away the remaining resist and gold layer. Then after removal of etching masks, another step of RIE can be performed to completely remove $MoS_2$ in the background and simultaneously thin down the thickness of final device. Using this method, we successfully fabricate $MoS_2$ waveguides with thickness of ~85 nm (figure S11c) from an original ~100 nm thick as-exfoliated flake (figure S11b).

Optical micrograph of fabricated devices is shown in figure S11c. The final thickness of the fabricated devices is measured with AFM scanning (inset in figure S11c). In order to avoid damaging the waveguides during the AFM scan we have also designed a small frame about 4 $\mu m$ thick around the fabricated devices made out of the same flake. As a result, we can perform AFM scanning on the frame to determine the thickness of the final device. In addition, we extract the surface roughness of as-exfoliated flake and after-fabricated devices from AFM scanning and find that the surface becomes rougher during the fabrication.

The average roughness is about 0.3 nm with maximum roughness of < 0.6 nm for as-exfoliated flake. However, the average roughness is almost 2 nm and the maximum roughness of 8 nm for fabricated devices. This increased top surface roughness contributes to additional optical scattering and absorption during the measurement. Nevertheless, our devices are smoother than previous demonstrations. The details of the fabricated devices are also examined with SEM imaging, Figure S12a. The SEM reveals that the fabricated waveguides have nearly straight side walls (due to a small thickness of devices) and rather smooth side wall surface (compared to prior fabrication examples [30]). The SEM images are also used to estimate the dimensions of the fabricated devices, including the waveguide footprint (determined to be ~320 nm). For example, due to raster scanning of the electron beam during lithographic patterning (inherently determined by the tool) structure dimensions in one of the directions appear to be slightly smaller than in the other (about 5%), see also discussion in figure S15. To further confirm the thickness of the waveguide, we have made

AFM topographic scan of one of the fabricated waveguides, Figure S12b and S12c. The AFM scanning and the thickness profiles above show that the width of the fabricated device is ~ 320 nm in width and ~ 85 nm in height. Moreover, the fabrication process, especially the reactive ion etching step, does not lead to a very tilted sidewall. Shown in figure S12c, the width at the bottom of structure is only slightly larger than the width at the top (30 – 40 nm wider, which is around 10%).

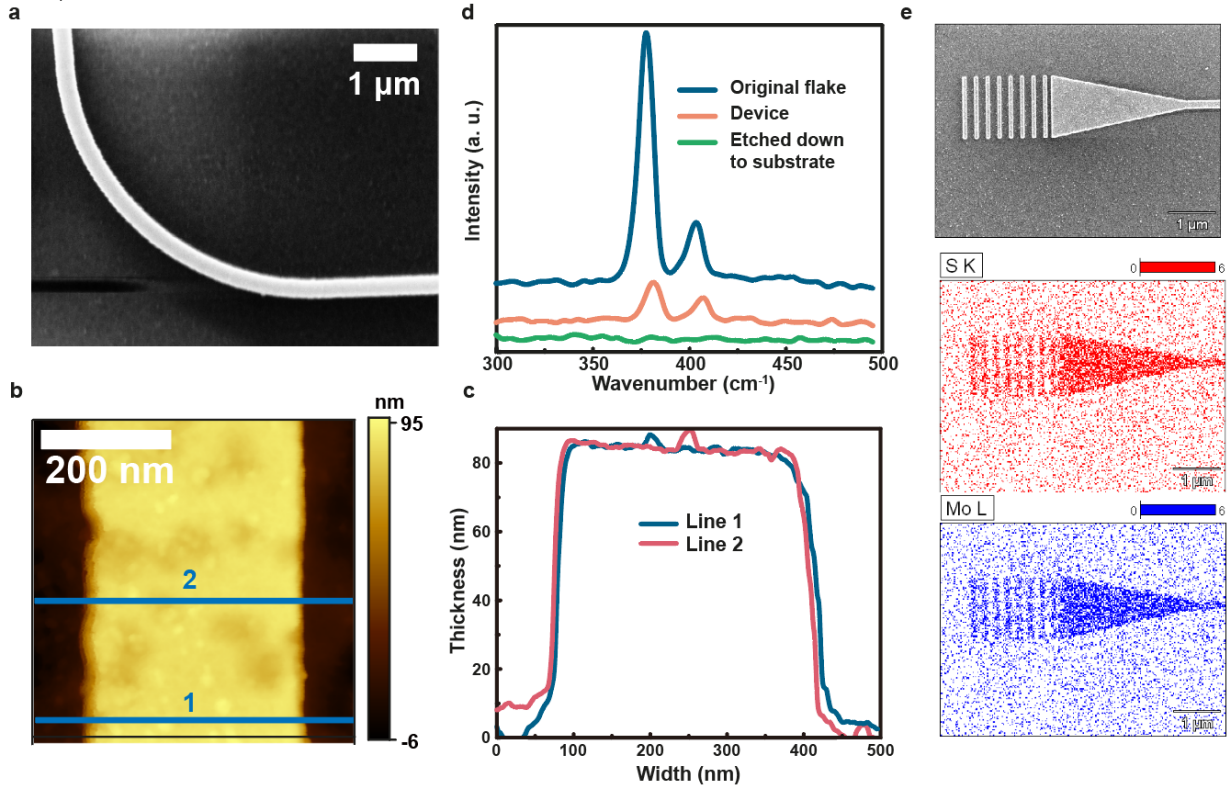

*Figure S12. Characterization of fabricated MoS$_2$ devices. **a,** Zoomed-in SEM image of the fabricated waveguide. **b,** AFM scanning of part of fabricated MoS$_2$ waveguide. **c,** Thickness profile of the waveguide along the two lines shown in **b**. **d,** Raman spectra of as-exfoliated flake, devices after fabrication and background area after fabrication. **e,** EDS mapping of fabricated MoS$_2$ grating coupler and taper.*

To confirm that the gold layer is completely removed, that the materials in the background are completely etched away, and the final devices are indeed made of MoS$_2$ we perform Raman spectroscopy and EDS. Figure S12d shows Raman spectra (obtained with Renishaw inVia™ confocal Raman microscope). Raman signal was measured from as-exfoliated flake (i.e., before e-beam lithography development) and after the final devices were made. The spot size of the Raman microscope is larger than the size of the fabricated devices, as a result the collected signal after fabrication is weaker in intensity. However, position of characteristic peaks is preserved and matches that of original flake. Figure S12e shows the EDS mapping of the fabricated waveguide.

## V. Far field measurements
### V.1. Experimental setup

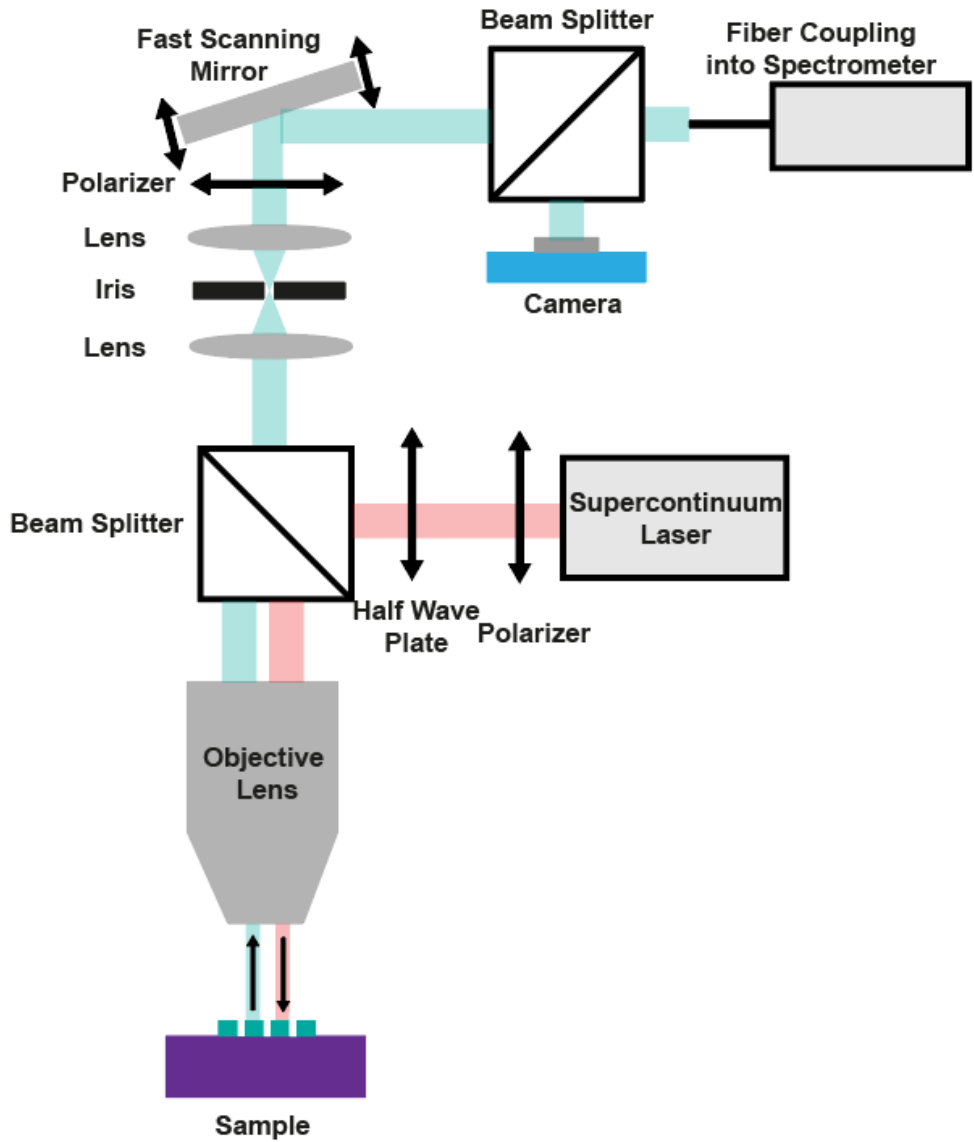

*Figure S13. Schematic illustration of the far-field measurement setup.*

Figure S13 shows the experimental setup used to probe the $MoS_2$ photonic chip. Light from a supercontinuum laser is used to probe the chip. A polarizer is used to ensure that the excitation light is polarized, and the half wave plate ensures constant power of laser at each polarization as it is rotated with respect to the grating. The light is guided to the sample through a 40x objective lens (Olympus LUCPlanFLN 0.6 NA). In our setup, the objective is fixed but the sample is on a XYZ translation stage. The output signal is collected through the same objective lens and transmitted to a camera (Point Grey Chameleon monochromatic CCD video camera part # CMLN-13S2M-CS) for imaging. To measure the spectrum of the transmitted light, we use a fast-scanning mirror (FSM-

CD300B) to couple the desired light from the output grating to a fiber (Thorlabs M64L01 10 μm core). We send a laser from the other side of the fiber to image on the same camera to identify the collection spot of the fiber. We then use the scanning mirror to ensure that the output grating light is on the collection spot. This way, the fiber itself acts as a pinhole. We then attach the output of the fiber to a detector (Princeton Instruments PIXIS CCD) and spectrometer (IsoPlane SCT-320 Imaging Spectrograph from Princeton Instruments) via the commercially available fiber attachment part to the spectrometer. We also use a polarizer in the output path. This output polarizer is perpendicular to the input polarizer, ensuring better rejection of the uncoupled input light.

### V.2. Measurement procedure

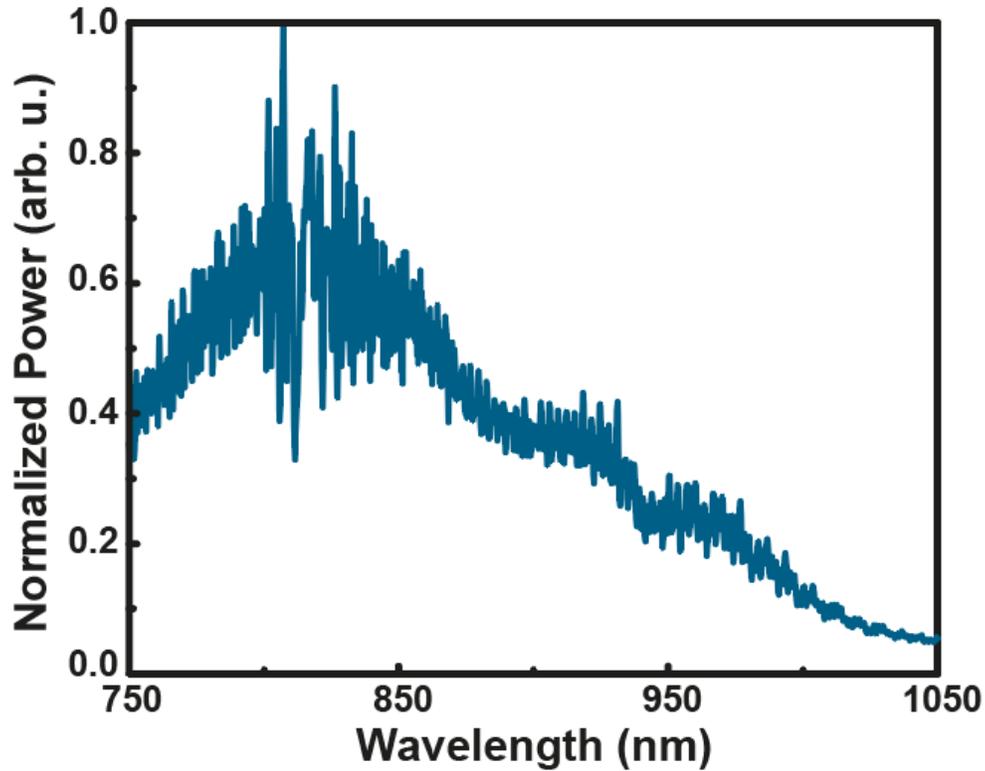

*Figure S14. Normalized reflected power of supercontinuum laser used in experiment from a silver mirror collected using setup shown in S13.*

The MoS$_2$ waveguides and other devices were measured using a confocal microscopy setup. A supercontinuum laser (Fianium WhiteLase Micro) is focused via an objective on one of the gratings (say, input grating). We collect the transmitted light through the same objective from the other grating (say, output grating). The microscope setup is designed to have both input and output gratings in the same field of view (estimated to be $\sim 40\ \mu m \times 30 \mu m$). We used a fiber-based pinhole in the confocal plane to collect the light from the output grating (as explained in the previous section).

To calibrate the spectral response of the whole setup (excluding the MoS$_2$ device), we place a silver mirror (Thorlabs UM10-AG) at the plane of the chip and measure the reflection (figure S14). To measure this reflection, we use a neutral density filter, which has flat spectral response over the wavelength of interest. We call this directly reflected light spectrum as the input spectrum $S_{in}(\lambda)$. The silver mirror has an average reflectance of 99.5% with negligible variation ($< 0.5\%$) across our wavelength of interest (700nm-1050nm). We note that, sensitivity of our spectrometer beyond 950nm is poor and hence the collected data in that range is noisy. Then we swap out the silver mirror with the photonic chip with MoS2 devices and measure the device transmission. The collected spectrum is termed as $S_{collected}(\lambda)$. The resulting MoS2 device spectra are then found as: $S_{out}(\lambda) = \frac{S_{collected}(\lambda)}{S_{in}(\lambda)}$.

## VI. Additional device simulations

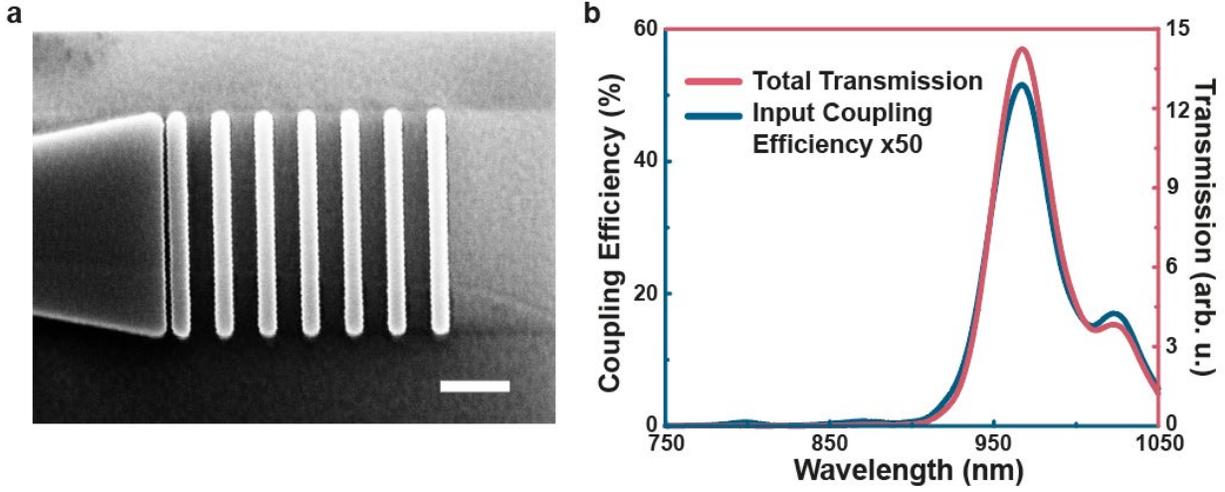

*Figure S15. **a**, Zoomed-in SEM image of a fabricated grating coupler. Scale bar: 1 μm. **b**, Simulated input coupling efficiency and total waveguide transmission profile for fabricated grating coupler.*

As we mentioned in Section II, because of fabrication imperfections, the geometric dimensions of the fabricated grating couplers differ from what we design originally (figure S8a). Instead of 160 nm, the width of fabricated grating bars is slightly wider (figure S15a). We carefully examine the zoomed-in SEM image and find that while the period of the grating is preserved at 600 nm, the duty cycle increases from 0.267 to 0.467 (each bar is 280 nm wide). As a result, it leads to a shift in the peak of the coupling efficiency. Based on this new grating parameters, we update FDTD calculations and plot the input coupler efficiency and total waveguide transmission profile in figure S15b. The total waveguide transmission profile has a good match with the experimental results (figure 2d in the main text).

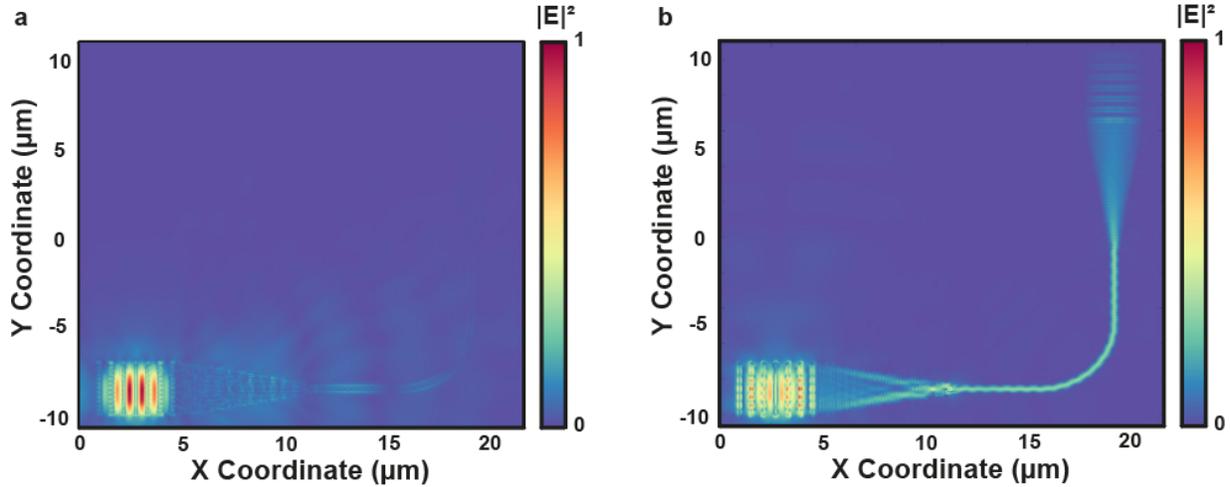

***Figure S16.*** *FDTD simulated mode profile of MoS$_2$ waveguide at **a** 800 nm and **b** 950 nm wavelengths.*

In section II, we indicate that the almost zero transmission for <900 nm is due to a strong material absorption of the waveguide combined with the low input coupling efficiency. To further study the role of optical extinction on excitation and guidance, we examine excited field in the waveguide as a function of the wavelength in figure S16. We plot excited field intensity at $\lambda = 800\ nm$ and $\lambda = 950\ nm$, respectively. As clearly seen, at 800 nm wavelength the scattering at the input coupler is very strong, indicating that only a small fraction of light is coupled into the waveguide. Combined with strong material absorption, only a very small fraction of optical power is guided to the output port at this wavelength. On the other hand, at 950 nm wavelength, the coupling efficiency is much higher and the material absorption is lower, allowing a large fraction of optical power to be transferred to the output coupler.

In Figure S17 we examine numerically cross polarized far-field measurement performance. Specifically, we plot the profiles of Ex and Ey electric field components at $\lambda = 950\ nm$ (field profiles are collected 1.5 $\mu m$ above the waveguide). The input grating (lower right corner of figure S17 a and b) is illuminated by Ey field, which excites TE mode in the waveguide. Collecting signal in Ex polarization allows minimizing scattering from the input grating coupler (figure S17a) so that the signal from the output grating is resolved, whereas in y-polarization light scattering from the input grating overshadows the output signal (figure S17b).

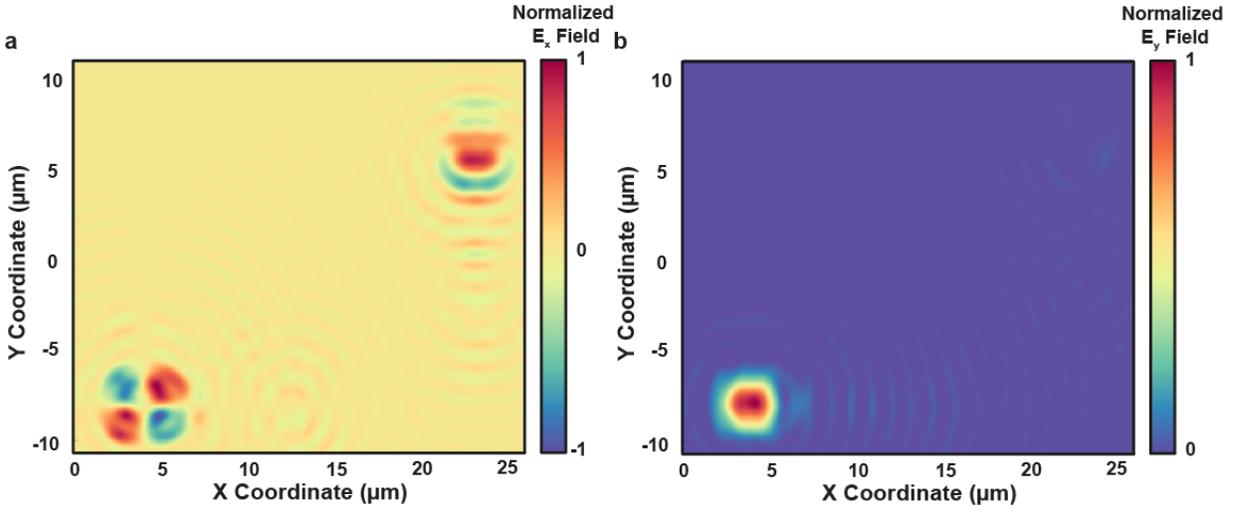

*Figure S17. Electric field profiles above MoS₂ device. a, $E_x$ and b, $E_y$ field profiles 1.5 μm above the device shown in figure S8a.*

## VII. Additional measurements
### VII.1 Symmetry study

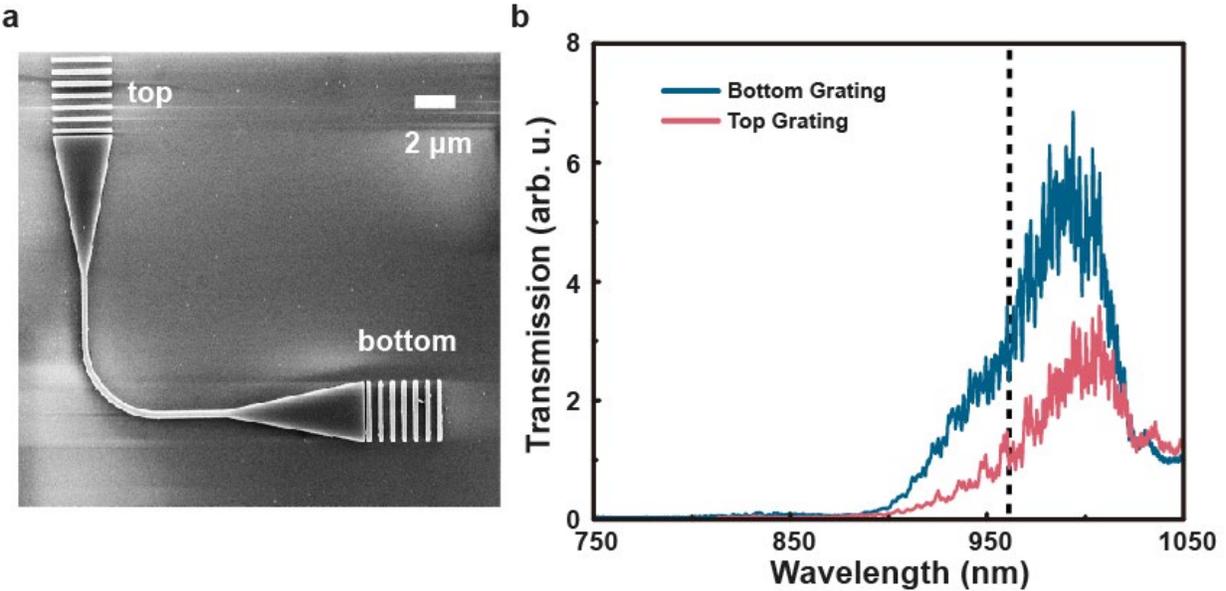

*Figure S18. Dependence on the choice of input grating. a, SEM image of a MoS₂ waveguide of 85 nm thickness with top and bottom grating couples labeled, respectively. b, Transmission spectra for light incident on bottom and the top gratings, respectively. The dashed line indicates the onset of bulk MoS₂ absorption band gap edge.*

In figure S18 we study dependence of device operation on the choice of input grating. Under ideal fabrication and measurement conditions one may expect that for identical gratings the device performance would be independent of the input grating choice. In our case, during the e-beam lithography writing, despite we design both the top and bottom gratings to be exactly identical, the features along *x* and *y* directions have different sizes, as shown in figure S18a. This difference leads to a slightly modified performance between the two gratings. Figure S18b plots transmission spectra for a device illuminated at the top and bottom grating couplers, respectively. A slight difference in transmission is seen.

## VII.2. Control device measurement

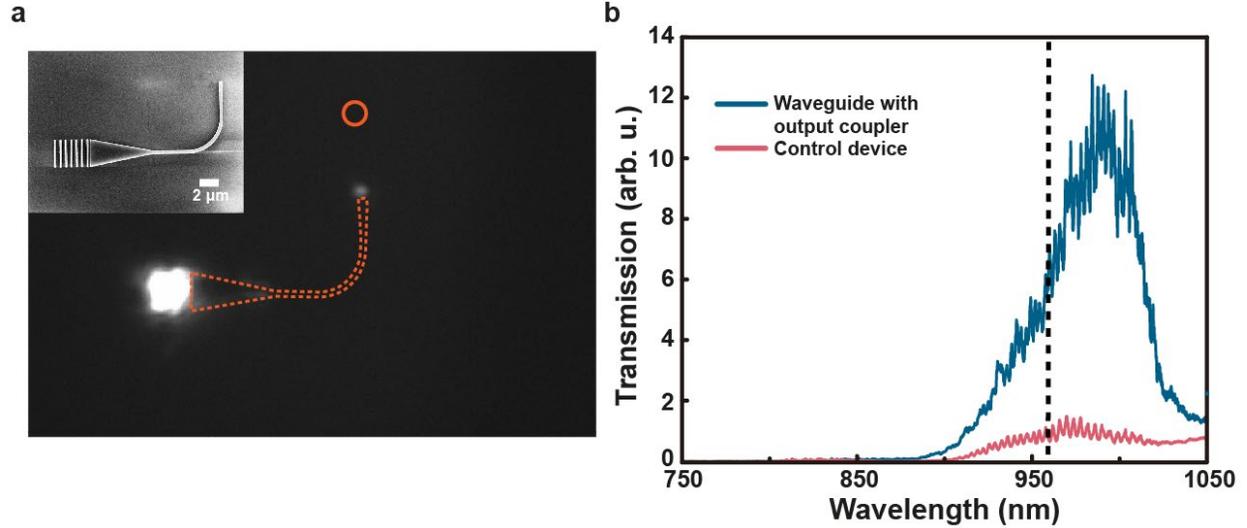

***Figure S19. Study of a control device. a,*** *Far-field image of $MoS_2$ control device that has input only grating coupler. Dashed contour denotes device profile. For reference, the circle marks the expected position of an output grating coupler. Inset: SEM image of the control device.* ***b,*** *Transmission spectra for an actual device (i.e., with input and output grating couplers) and control device (i.e., without output grating coupler), respectively. Dashed line indicates the onset of bulk $MoS_2$ absorption band gap edge.*

To ensure that light collected at the position of the output grating is not mere scattering, but is light guided by the device, we fabricate a set of control devices (figure S19a inset). For our control devices, we deliberately design only input grating coupler with no output grating coupler. Comparison of transmission spectra with output grating coupler (regular device) and without output grating coupler (i.e., control device) is shown in figure S19b. It is noteworthy that the transmitted signal from control device is not collected at the circled area – where the output coupler is supposed to be, but at the bright spot at the end of waveguide, which is generated by the scattering of light. The data measured from the encircled area in figure S19a (i.e., where grating coupler is supposed to be) is below the noise level, suggesting high spatial contrast of our measurement setup.

## VII.3. Polarization dependent studies

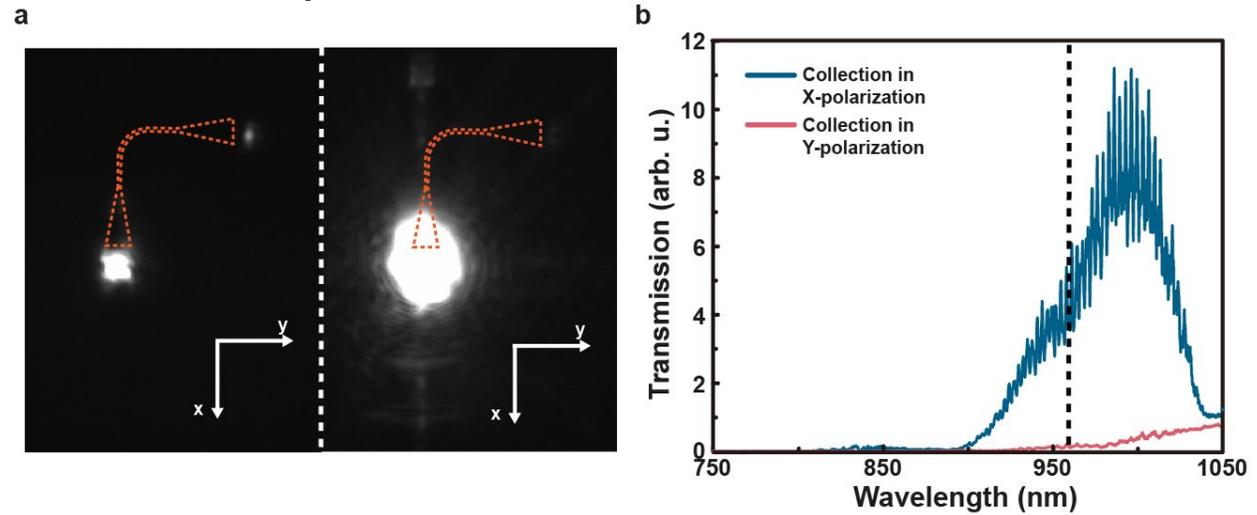

*Figure S20. Polarization dependent studies of the device. **a**, Optical micrograph of the scattered light measured in x-polarization (left) and y-polarization (right), i.e., orthogonal and parallel to the to the incident beam polarization, respectively. Polarization of the incident beam is along y-axis. The contour of the measured waveguide is denoted with a dashed curve. **b**, Transmission spectra in x and y polarizations. Dashed line indicates the onset of bulk $MoS_2$ absorption band gap edge.*

As shown in figure S8a, we introduce a 90° bending for cross polarization far field measurement. In our cross polarization measurements, in the output optical path we introduce a linear polarizer perpendicular to the excitation polarization. In figure S20 we examine the far field imaging and spectra collection in polarization parallel ($y$) and perpendicular ($x$) to the incident beam polarization (polarized along the $y$-axis). Figure S20a shows optical images of the scattered light in $x$ and $y$ polarizations, respectively. The output emission spot is clearly visible in the $x$ polarization. Furthermore, light collection in polarization orthogonal to that of the incident beam allows filtering our strong scattering at the input grating coupler. These experimental data correlate well with simulations shown in figure S17. Figure S20b shows spectra collected at the output grating in $x$ and $y$ polarizations. Signal collected in the $y$ polarization is close to the noise floor level.

Our waveguides are designed to support only a fundamental TE mode (i.e., predominantly polarized along the $MoS_2$ crystallographic plane). To verify that only TE modes can be excited and guided in our device we have studied polarization dependent excitation of the device. Figure S21a shows optical micrographs in cross polarized light for waveguide excitation along the $x$ axis (i.e., perpendicular to the grating fingers) and along the $y$ axis (i.e., parallel to the grating fingers). Evidently, guided waves are excited only in the case of $y$-polarized incident beam, at which case guided TE waves are excited. Excitation of TM waves is not observed. Figure S21b shows the measured transmission spectra for the two cases, i.e., $x$ and $y$ polarized incident beams (in both cases cross-polarized output is assumed). As expected, nearly no signal is guided for incident beam polarized along the $x$ axis.

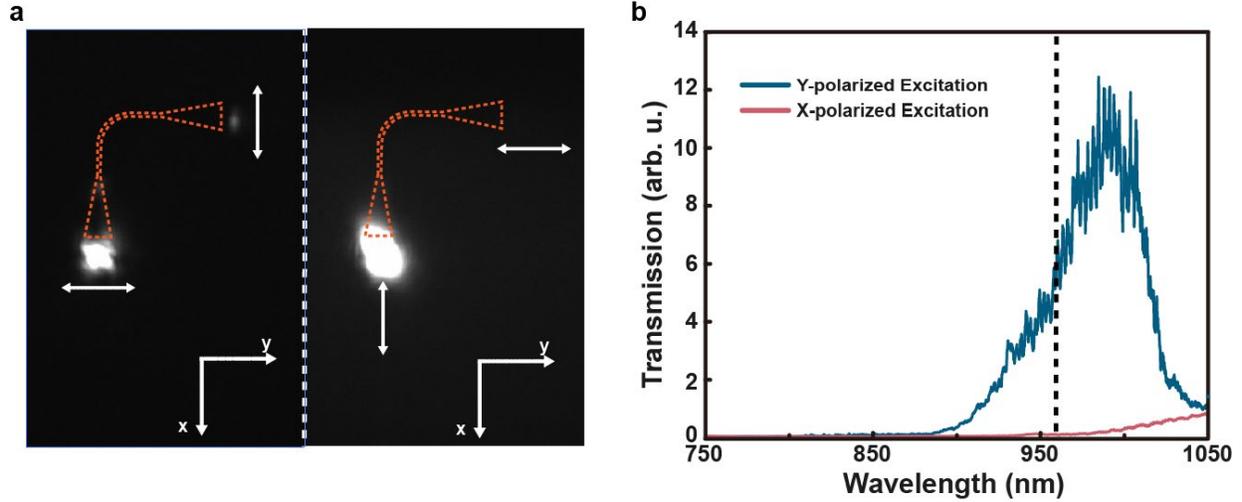

*Figure S21. Study of TE and TM wave excitation. **a,** Optical images of collected signal in cross-polarized light for y-polarized illumination (left) and x-polarized illumination (right). The arrows indicate the polarization at the input and output, respectively. **b,** Cross polarized transmission spectra for x and y incident polarizations, respectively. Dashed line indicates the onset of bulk MoS$_2$ absorption band gap edge.*

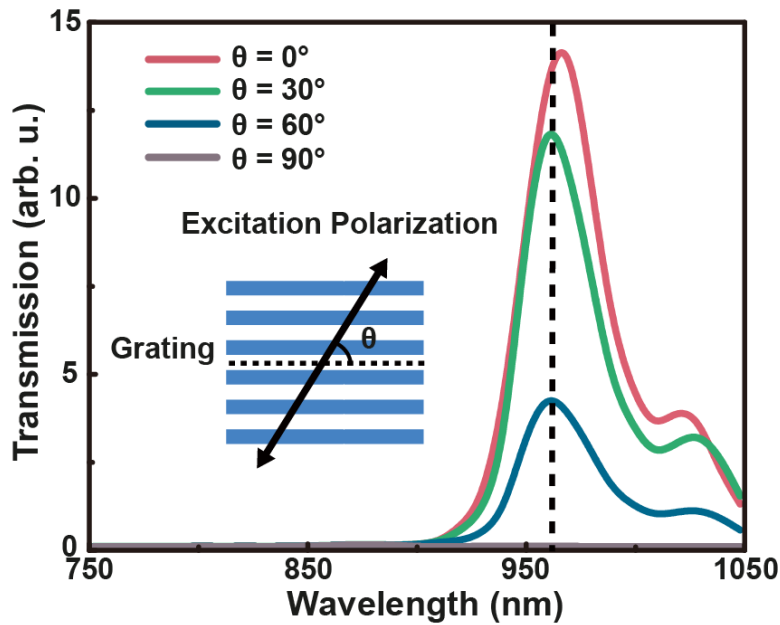

*Figure S22. Theoretical study of polarization dependent transmission using FDTD simulation. The dashed line indicates the onset of bulk MoS$_2$ absorption band gap edge.*

Finally, in figure S22 we plot calculated polarization dependent transmission. Here we assume simulation parameters similar to those in Sec. II and Sec. VI. We vary the incident beam polarization and study the signal transmission (output beam polarization is along the direction of the output grating coupler). The resulting transmission spectra are shown in figure S22. As expected, transmission decreases as the incident beam polarization is rotated and nearly vanishes for $\theta = 90°$. At this polarization, i.e., $\theta = 90°$, TE modes are practically not excited in the waveguide. This simulation agrees well with the experimental spectra shown in figure 2e in main text.

## VIII. Near field measurements.

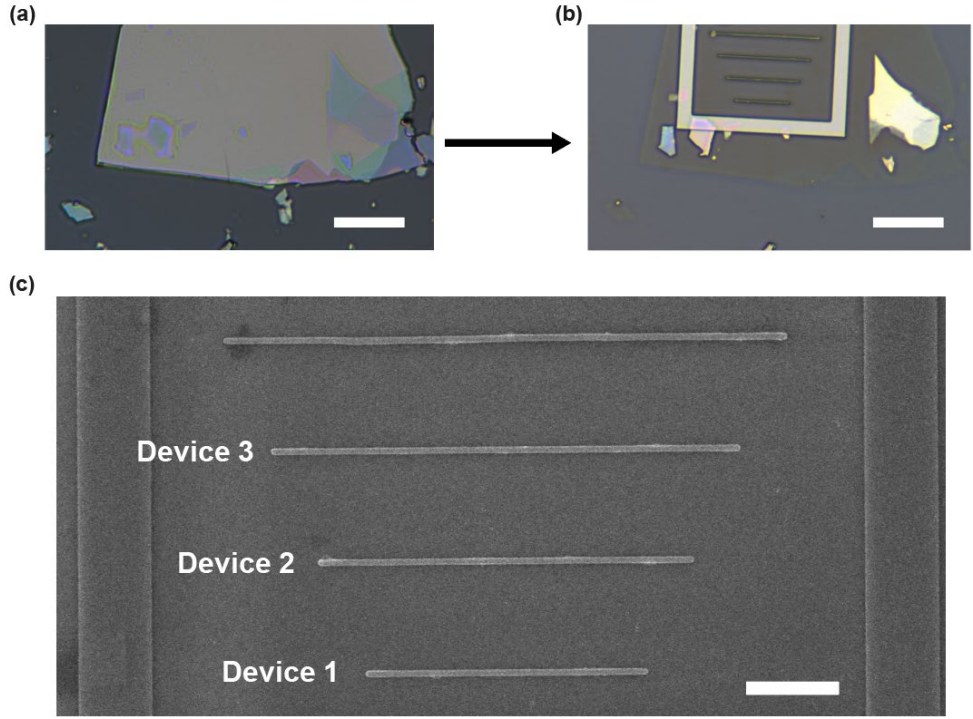

***Figure S23. Device fabrication for near-field measurements***. ***a,*** *Microscope optical image of the as-exfoliated MoS$_2$ flake on Si/SiO$_2$ substrate. Scale bar: 20 μm.* ***b,*** *Microscope optical image of a post-fabrication flake. Scale bar: 20 μm.* ***c,*** *SEM image of the fabricated devices. Scale bar: 5 μm.*

We start fabrication of deeply subwavelength MoS$_2$ ridge waveguides for near-field imaging by exfoliating a large area flake of ~85 nm thickness atop the SiO$_2$/Si substrate, as shown in figure S23a. After the fabrication processes described in section IV, we achieve four different devices with different lengths but the identical thickness of ~ 80 nm (figure S23b). The AFM of the device can be found in figure 3b in the main text. Figure S23c shows the SEM image of the fabricated devices. Devices 1, 2, and 3 have lengths of 15 μm, 20 μm, and 25 μm, respectively, and the top device is 30 μm long. The s-SNOM scanning results of device 1 at different wavelengths are shown in the main text. To show that our results are reproducible, in figure S24, we plot s-SNOM images for devices 2 and 3. Dynamics similar to that of device 1 is seen. In all cases, we do not observe wave propagation below 900nm.

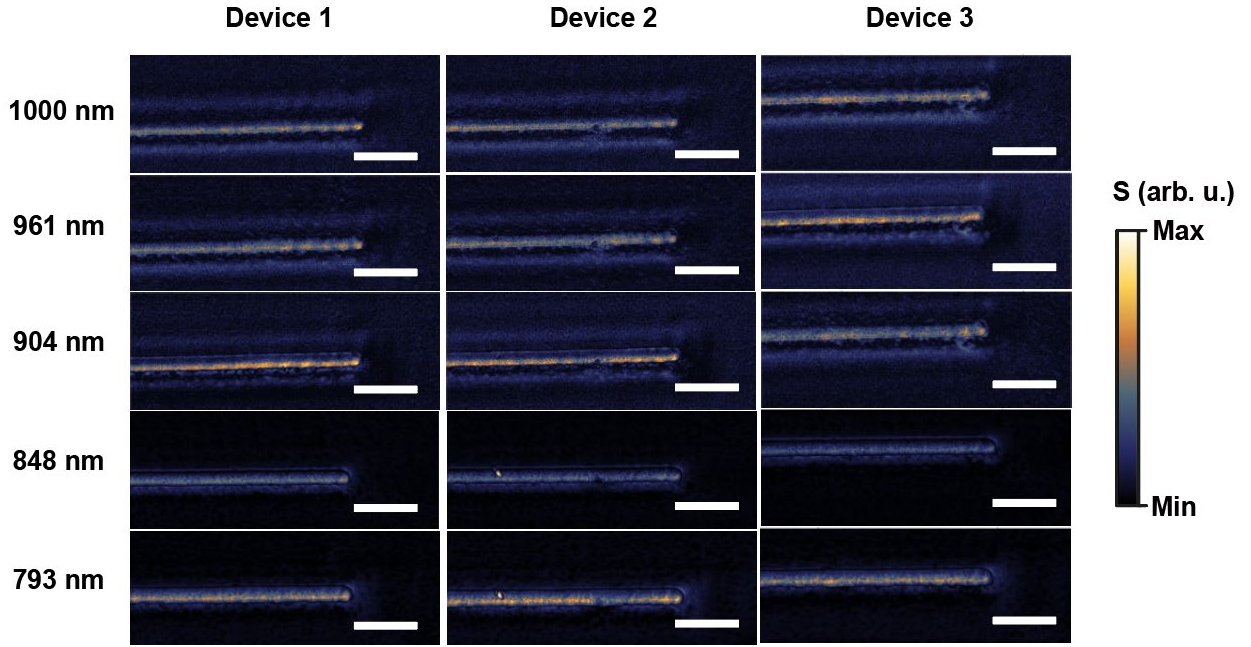

***Figure S24.*** *s-SNOM images at the illumination wavelength from 1000 nm to 793 nm of the devices shown in figure S23. Scaler bar: 2 μm.*

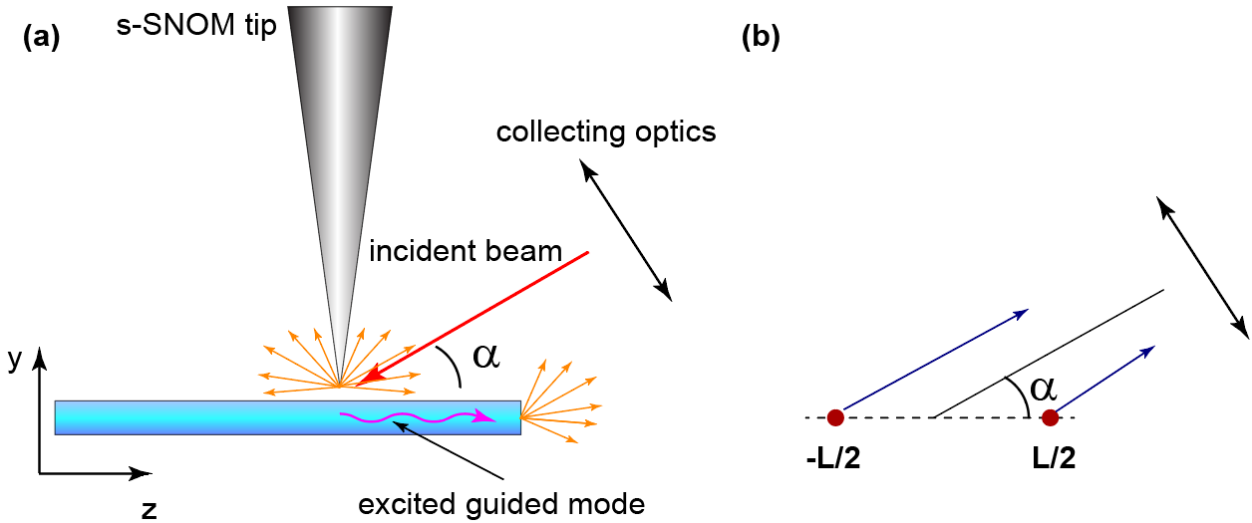

***Figure S25***. ***a,*** *Schematic of s-SNOM imaging setup.* ***b,*** *Equivalent two-dipole linear array picture.*

**VIII.1. s-SNOM imaging mechanism and the extraction of wavelength and damping**

In our NIR s-SNOM measurements, the s-SNOM tip is illuminated by a NIR laser incident at an angle $\alpha$ (figure S25a). Upon tip illumination, a TE exciton-dressed waveguide mode is excited, which propagates from the position of the tip to the edge of the TMDC waveguide. Reaching the waveguide edge, the mode is partially scattered into the free space. The observed interference pattern is then formed by the waves scattered directly off the tip and guided waves scattered at the waveguide edge (see figure S25a). For an *s*-polarized incident beam scattered fields can be approximated as dipolar fields with dipole polarization along the *x* axis (i.e., along the direction of incident electric field polarization). Such dipolar approximation is justified as long as the tip and waveguide cross-sections are smaller than the incident wavelength and the waveguide supports only a single TE mode. These approximations hold well in our experiments.

For two dipoles with a distance $L$ between them, the radiation pattern can be found based on far-field approximation. In this case, the collected field is described by a two element linear phased array, figure S25b. In this approximation, we can express the overall collected far-field as a superposition of two dipoles (tip dipole and waveguide edge dipole):

$$\boldsymbol{E} \simeq a_1 e^{i\psi_1} \boldsymbol{E}_0 e^{-\frac{i\tilde{\kappa}_0 \cos(\alpha)L}{2}} + a_2 e^{i\psi_2} \boldsymbol{E}_0 e^{i\text{Re}(\tilde{\kappa})L} e^{-\gamma L} e^{\frac{i\tilde{\kappa}_0 \cos(\alpha)L}{2}}, \quad (1)$$

here the first term corresponds to dipole excited at the tip, $a_1$ and $\psi_1$ are respective amplitude and phase, $\boldsymbol{E}_0$ is the incident field at the tip, $\tilde{\kappa}_0 = \frac{2\pi}{\lambda_0}$ is the free space wavevector, $\lambda_0$ is the wavelength of the incident laser beam, $\alpha$ is the collection beam path angle, which in our case coincides with the angle of incidence (figure S25a and S25b). The second term in Eq. (1) corresponds to the dipole excited at the waveguide edge. In this case, the field that excites the edge dipole is that of the excited waveguide mode that propagates distance $L$. Hence, phase $e^{i\text{Re}(\tilde{\kappa})L}$ and amplitude decay $e^{-\gamma L}$ are introduced, where $\tilde{\kappa}$ is the guided mode wavevector and $\gamma = \text{Im}(\tilde{\kappa})$ denotes spatial decay rate.

The observed interference fringe pattern is then given by $|\boldsymbol{E}|^2$, which, in turn, is dominated by the term $\frac{1}{2} a_1 a_2 |\boldsymbol{E}_0|^2 e^{-2\gamma L} \cos(\tilde{\kappa}_0 \cos(\alpha) L - \text{Re}(\tilde{\kappa})L + \Delta\psi)$. Assuming that the incident field intensity, $|\boldsymbol{E}_0|^2$, amplitudes, $a_1$ and $a_2$, and phase difference, $\Delta\psi = \psi_1 - \psi_2$, do not change with tip position, we can relate experimentally observed interference fringe period, $\rho$, with phase $\tilde{\kappa}_0 \cos(\alpha) L - \text{Re}(\tilde{\kappa})L$ as follows:

$$\rho = \frac{2\pi}{[\tilde{\kappa}_0 \cos(\alpha) - \text{Re}(\tilde{\kappa})]}. \quad (2)$$

Using this expression, we extract the wavelength of the guided TE exciton-dressed mode:

$$\rho = \frac{1}{\frac{1}{\lambda_0}\cos(\alpha) - \frac{1}{\lambda}} = \frac{\lambda_0 \lambda}{\lambda \cos(\alpha) - \lambda_0}, \quad (3)$$

here $\lambda$ is the wavelength of the guided TE exciton-dressed mode. Expression (3) was used to plot figure 3i of the main text.

In addition, by tracing the decay of the interference intensity as a function of tip position, spatial decay rate $\gamma$ can be found. The propagation length is then given as:

$$L_p = \frac{1}{2\gamma} \quad (4)$$

# IX. Y-junction beam splitter and Mach-Zehnder interferometer.

## IX. 1. Device fabrication and characterization

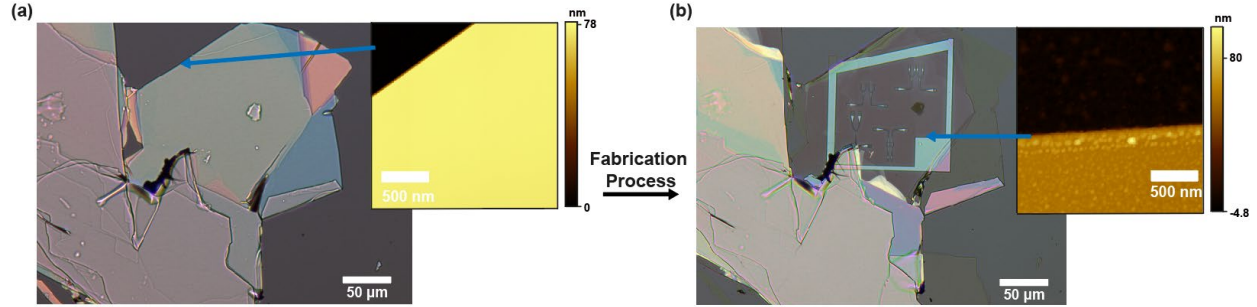

*Figure S26. Device fabrication. **a,** Microscope optical image and AFM scanning of the as-exfoliated $MoS_2$ flake on Si/SiO$_2$ substrate. **b,** Microscope optical image and AFM scanning of post-fabricated flake. The AFM scans in both **a** and **b** are performed at the locations indicted by the arrows.*

We start fabrication of deeply subwavelength $MoS_2$ Y-splitters and Mach-Zehnder interferometers by exfoliating a large area flake of ~75 nm thickness atop SiO2/Si substrate, as shown in figure S26a. The average roughness of the as-exfoliated flake extracted from AFM scan is 0.05 nm and the maximum roughness is 0.41 nm. After fabrication processes described in section IV, we achieve several Y-splitter devices and Mach-Zehnder interferometers with a thickness of ~60 nm. We again reserve 4 μm thick frame made out of $MoS_2$ around the devices for AFM measurements (figure S26b). The average roughness of the fabricated devices is 1.4 nm and the maximum roughness height is 7.8 nm. The increased top surface roughness of the device results in additional optical scattering and extinction.

## IX. 2. Device simulations

In figure S27 we plot profile of $E_x$ component of the electric field. Here y-polarized light excitation is assumed. Two distinct emission spots at the output are observed. Calculations show 50/50 power splitting performance.

Finally, in figure 28 we study theoretically the performance of symmetric and antisymmetric Mach-Zehnder interferometers. In this calculation we do not take into account input and output grating couplers. Specifically, we excite the devices with a fundamental TE mode from one end of the device and examine transmitted power at the other end. Respective transmission spectra are plotted in figure S28. For a symmetric Mach-Zehnder interferometer contractive interference leads to signal transmission through the device. At the same time transmission is affected by optical extinction. As a result, for shorter wavelengths, where optical losses are higher, transmitted signal power is lower. For the antisymmetric Mach-Zehnder interferometer destructive interference is expected. However, due to a difference in the length of interferometer arms the insertion loss is different. As a result, complete destructive interference is not observed. At the same time, calculations predict significantly lower transmission for an asymmetric device. Predicted difference is comparable to that observed experimentally.

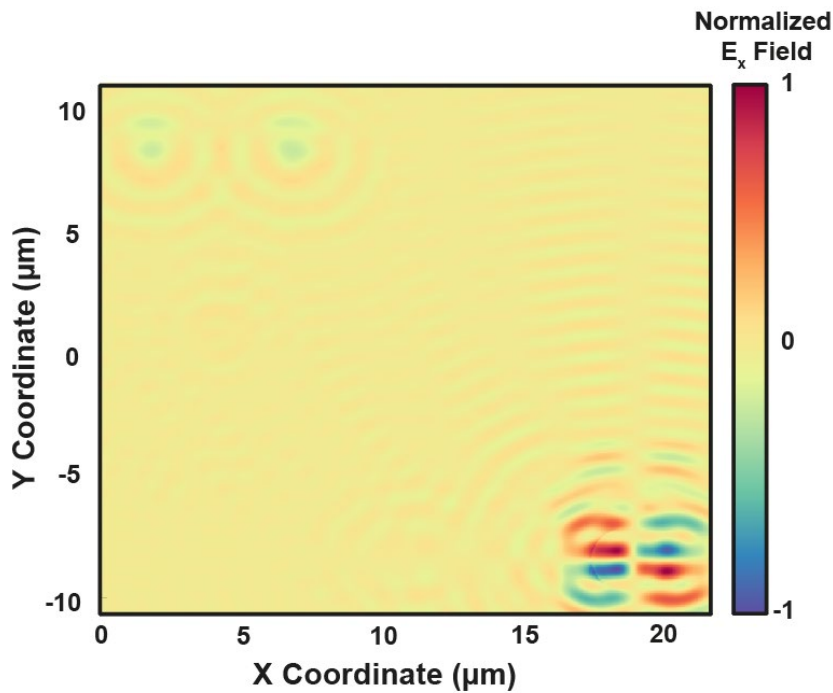

*Figure S27. Y-junction excitation.* Ex electric field profile above the device at $\lambda = 950\ nm$.

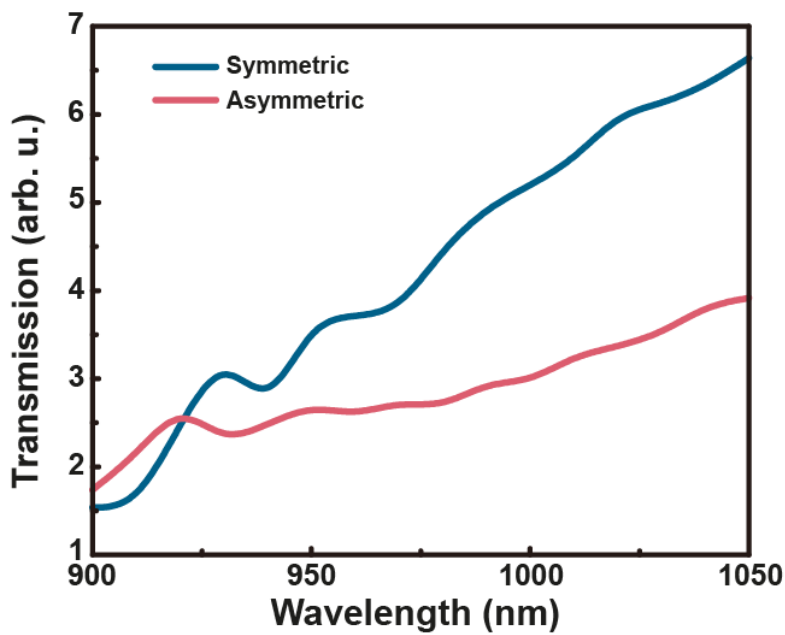

*Figure S28. Study of Mach-Zehnder interferometer.* Transmission of symmetric and asymmetric Mach-Zehnder interferometers.